\newif\ifsingle
\singletrue 

\ifsingle
\documentclass[11pt,draftclsnofoot, onecolumn]{IEEEtran}		
\else		
\documentclass[10pt,final, twocolumn]{IEEEtran}
\fi

\usepackage{times}
\usepackage{amsmath,dsfont}
\usepackage{amssymb,amsthm}
\usepackage{epsfig,verbatim}
\usepackage{subfigure}
\usepackage{setspace}
\usepackage{color}
\usepackage{cite}
\usepackage{epstopdf}
\usepackage{graphics}
\usepackage{accents}
\usepackage{acronym}
\usepackage[bookmarks,colorlinks]{hyperref}
\usepackage{booktabs}
\usepackage{mathtools}
\usepackage{algorithm}
\usepackage{algorithmic}
\usepackage{enumitem}
\usepackage{pstool}
\usepackage{tabularx}

\ifsingle

\else

\fi 

\acrodef{adc}[ADC]{Analog-to-Digital Convertor}
\acrodef{dac}[DAC]{digital-to-analog convertor}
\acrodef{cs}[CS]{Compressed Sensing}
\acrodef{dtft}[DTFT]{discrete-time Fourier transform}
\acrodef{dnn}[DNN]{deep neural network}
\acrodef{csi}[CSI]{channel state information}
\acrodef{map}[MAP]{maximum a-posteriori probability}
\acrodef{snr}[SNR]{signal-to-noise ratio}
\acrodef{sinr}[SINR]{signal-to-interference-and-noise ratio}
\acrodef{bs}[BS]{Base Station}
\acrodef{iot}[IOT]{Interent of Things}
\acrodef{mimo}[MIMO]{Multiple-Input Multiple-Output}
\acrodef{mse}[MSE]{mean-squared error}
\acrodef{pdf}[PDF]{probability density function}
\acrodef{rv}[RV]{random variable}
\acrodef{fec}[FEC]{forward error correction}
\acrodef{rs}[RS]{Reed-Solomon}
\acrodef{lti}[LTI]{linear time-invariant}
\acrodef{wss}[WSS]{wide-sense stationary}
\acrodef{psd}[PSD]{power spectral density}
\acrodef{ser}[SER]{symbol error rate}
\acrodef{ber}[BER]{bit error rate}
\acrodef{isi}[ISI]{intersymbol interference}
\acrodef{awgn}[AWGN]{additive white Gaussian noise}
\acrodef{ut}[UTs]{User Terminals}
\acrodef{mmw}[mmWave]{millimeter wave}
\acrodef{ris}[RIS]{reconfigurable intelligent surface}
\acrodef{dma}[DMA]{Dynamic Metasurface Antenna}
\acrodef{5G}{fifth generation}

\IEEEoverridecommandlockouts

\title{Near-Field Sparse Channel Representation and Estimation in 6G Wireless Communications}
\author{
Xing Zhang,~\IEEEmembership{Member,~IEEE}, Haiyang Zhang,~\IEEEmembership{Member,~IEEE}, and Yonina C. Eldar, ~\IEEEmembership{Fellow,~IEEE}
}

\vspace{-0.75cm}

\begin{document}
	
	\maketitle
	\pagestyle{plain}
	\thispagestyle{plain}
	


\begin{abstract}
The employment of extremely large antenna arrays and high-frequency signaling makes future 6G wireless communications likely to operate in the near-field region. In this case, the spherical wave assumption which takes into account both the user angle and distance is more accurate than the conventional planar one that is only related to the user angle. Therefore, the conventional planar wave based far-field channel model as well as its associated estimation algorithms needs to be reconsidered. Here we first propose a distance-parameterized angular-domain sparse model to represent the near-field channel.
In this model, the user distance is included in the dictionary as an unknown parameter, so that the number of dictionary columns depends only on the angular space division. This is different from the existing polar-domain near-field channel model where the dictionary is constructed on an angle-distance two-dimensional (2D) space. Next,
based on this model, joint dictionary learning and sparse recovery based channel estimation methods are proposed for both line of sight (LoS) and multi-path settings. To further demonstrate the effectiveness of the suggested algorithms, recovery conditions and computational complexity are 
studied. Our analysis shows that
with the decrease of distance estimation error in the dictionary, the angular-domain sparse vector can be exactly recovered after a few iterations. The high storage burden and dictionary coherence issues that arise in the polar-domain 2D representation are well addressed.
Finally,
simulations in multi-user communication scenarios support the superiority of the proposed near-field channel sparse representation and estimation over the existing polar-domain method in channel estimation error.

\end{abstract}

\vspace{-0.2cm}
\section{Introduction}
With the commercialization of 5G, the next sixth-generation (6G) wireless communication, which is required to achieve very high data rates, i.e., up to 1 Tb/s, has become a research hotspot of both industry and academia \cite{rajatheva2020white,chowdhury20206g,letaief2019roadmap}. Two key physical-layer technologies to support the ever-increasing data rate are the employment of extremely large antenna arrays with hundreds or even thousands of radiating elements, and high-frequency signaling at millimeter-wave (mmWave) or sub-terahertz (THz) bands \cite{shlezinger2021dynamic,de2020non,petrov2020ieee}. The former can significantly improve spectral efficiency while the latter provides large available bandwidth. However, the use of extremely large antenna arrays and high-frequency signaling results in fundamental changes in the characteristics of electromagnetic wave propagation.

With the increase of the antenna aperture and decrease of signal wavelength, future 6G wireless communications are likely to take place in the radiating near-filed (Fresnel) region, different from conventional wireless systems typically operating in the far-field region \cite{zhang20226g,zhang2022near,cui2022near}. The boundary between the two is the Fraunhofer limit, defined as $d_F=\frac{2D^2}{\lambda}$, where $D$ is the antenna diameter and $\lambda$ the wavelength \cite{nepa2017near}. Consider the case where a base station (BS) is equipped with a $D=1$ meter uniform linear array (ULA) and the carrier frequency is $30$ GHz; the corresponding radiating near-field region extends up to $200$ meters. Consequently, in 6G networks, most users are likely to be located in the near-field area of the BS.
In this case, the conventional planar wavefront assumption adopted in far-field communication is not valid. Instead, in near-field communication, the spherical wavefront assumption is more accurate as the time difference of arrival between each antenna depends on both the angles of arrival and the distance \cite{zhang20226g,cui2022channel}. This property brings forth new opportunities for future wireless communication systems design. For example, spherical wavefronts have been exploited to generate near-field beam focusing that allows focusing signals on a specific spatial region, so that interference from other users in the same direction can be mitigated \cite{zhang2022beam}. In 6G-enabled internet of everything mobile networks, beam focusing can be applied to reduce energy pollution in wireless power transfer \cite{zhang2022near}.
However, the nature of the spherical wavefront poses significant challenges to signal processing for 6G, since both the model and algorithms under the planar wavefront case need to be reconsidered.

One interesting aspect is the near-field sparse channel  representation and estimation. Accurate acquisition of channel state information is the basis for signal detection at the receiver, as well as some preprocessing at the transmitter, e.g., the aforementioned near-field beam focusing. For channel estimation in large-scale antenna high-frequency  systems, sparsity has long been exploited to reduce pilot overhead \cite{xiong2016channel,cheng2020channel}.
In the current 5G massive multiple-input multiple-output (MIMO) communication mmWave system, which still operates in the far-field region, the steering vector of the antenna array is only related to the angle of arrival. Therefore, by exploiting the path sparsity in the angular-domain, the channel can be represented by a Fourier dictionary-based sparse model. Based on this model, various sparse recovery algorithms have been proposed to address the far-field sparse channel estimation problem.
For example, in \cite{Lee2016channel}, based on the angular-domain Fourier dictionary, the orthogonal matching pursuit (OMP) algorithm was utilized for channel estimation of hybrid massive MIMO systems. In \cite{Bellili2019generalized}, generalized approximate message passing was employed for massive MIMO mmWave channel estimation, and in \cite{Ding2018Bayesian}, in the presence of quantization noise, the angular-domain sparse channel estimation was formulated into a sparse Bayesian learning problem and solved by the variational Bayesian method.

In future 6G systems of which a major portion will take place in the radiating near-field, both the channel sparse representation model and its associated sparse estimation algorithm need to be reconsidered. Under the spherical wavefront assumption, the array steering vector depends on both the angles of arrival and the user distance. As a result, the Fourier dictionary which only considers the angular-domain is not appropriate to model the near-field channel. In \cite{cui2022channel}, a polar-domain representation method was proposed, where the dictionary is constructed by sampling both the angle space and the distance range. This approach, however, suffers from high storage burden and high computational complexity as it is an angle-distance 2D dictionary. Moreover, to get finer resolution, the angle and distance sampling interval should be as small as possible, resulting in high coherence between columns. To mitigate these issues, a nonuniform distance sampling method was further proposed \cite{cui2022channel}, allowing dense sampling in near distances and sparse sampling in far distances. Besides high complexity in dictionary construction, the nonuniform distance sampling method also suffers from performance loss, i.e., distance estimation accuracy. The polar-domain representation method was further applied to channel modeling and estimation in hybrid-field settings where the scatters can be in far-field or near-field \cite{wei2021channel}.

In this work, we propose a distance-parameterized angular-domain sparse representation model for the near-field channel, together with a joint dictionary learning and sparse recovery based channel estimation algorithm, namely, dictionary learning orthogonal  matching pursuit (DL-OMP). Specifically, the main contributions of this work are as follows:

\begin{itemize}
\item
By exploiting the fact that with given observations, the angle and distance are coupled, a distance-parameterized angular-domain sparse near-field channel representation model is proposed and analyzed in detail. It is different from the polar-domain representation as the size of the constructed dictionary only depends on the angular resolution. This allows to overcome challenges that arise from the polar-domain method such as the storage burden and high column coherence.
\item
Based on the distance-parameterized angular-domain dictionary, for both line of sight (LoS) and multi-path settings, joint dictionary learning and sparse recovery based near-field channel estimation algorithms are provided. The proposed DL-OMPs iteratively estimate the sparse vector and update the dictionary, achieving accurate estimations of both the angle of arrival and the distance for each path. These parameters can not only be utilized to reconstruct the channel vector, but also provide user location information.
\item
Theoretical analysis of the algorithms is provided, including the restricted isometry property (RIP)-based recovery condition, and the
storage and computational complexity. The recovery condition
indicates that with the decrease of the distance estimation error in the dictionary, the angular-domain sparse vector can be exactly recovered after a few iterations.
The storage burden is determined by the size of the constructed dictionary. As the size of the distance-parameterized angular-domain dictionary only depends on the angular resolution, it is evident that the proposed representation requires lower storage compared with the polar-domain 2D dictionary, especially when the near-field range is large. The smaller size also makes the proposed algorithm less computational complex. Furthermore, simulations in multi-user communication settings show that the proposed near-field sparse channel representation and estimation outperform the existing polar-domain method in channel estimation normalized mean square error (NMSE). This can be explained by the low dictionary coherence of the proposed method, which facilitates more accurate sparse recovery.
\end{itemize}

The rest of this paper is organized as follows. In Section \ref{sec:prelimi}, brief background on the channel models for both far-field and near-field massive MIMO communication systems is presented. Next, in Section \ref{sec:CE_Alg}, a distance-parameterized angular-domain sparse representation model is provided, followed by joint dictionary learning and sparse recovery-based near-field channel estimation algorithms for both LoS and multi-path channels. Discussions on the recovery condition, storage and computational complexity of the proposed algorithm are also provided.
In Section \ref{sec:Sim}, numerical simulations demonstrate the superiority of the proposed DL-OMP algorithm in near-field channel estimation NMSEs over the polar-domain method. Section \ref{sec:Conc} concludes the paper.

\textit{Notation:} Scalar quantities, column vectors and matrices are denoted by lowercase letters, $a$, bold lowercase letters, $\textbf{a}$, and bold uppercase letters, $\textbf{A}$, respectively. The superscripts ${{(\cdot)}^{T}}$, ${{(\cdot)}^{H}}$ and ${{(\cdot)}^{\dag}}$ are transpose, Hermitian transpose and Moore-Penrose pseudo-inverse operators. The symbol $|\cdot|$ denotes the absolute value of a scalar, $||\cdot||$ denotes the norm of a vector, and $<\cdot,\cdot>$ is the inner product in Euclidean space.
The symbol $\text{diag}(\textbf{a})$ represents a diagonal matrix whose diagonal elements are elements of the vector $\textbf{a}$.
\section{Preliminaries on Channel Models}
\label{sec:prelimi}
In this section, we provide some preliminaries on the channel models in both far-field and near-field massive multiple-input multiple-output (MIMO) communications. In particular, a general channel expression based on free-space electromagnetic wave propagation theory is introduced. Its simplified representation in far-field based on the planar wavefront assumption is provided in Section \ref{subsec:FFCM}, and in near-field based on the spherical wavefront assumption is provided in Section \ref{subsec:NFCM}.

Consider an uplink multiuser massive MIMO communication system where the base station (BS) is equipped with an $N$ antenna uniform linear array (ULA). The antenna spacing is $d=\frac{\lambda_c}{2}$, with $\lambda_c$ denoting the carrier wavelength. The coordinate of the $n$th BS antenna is $(0,(n-1)d), n=1,\ldots, N$. For uplink channel estimation, we assume different users transmit mutual orthogonal pilot sequences to the BS, so that the channel estimation for each user can be performed independently \cite{bjornson2016massive}. Without loss of generality, we consider a single-antenna user located at $(x,y)$, with polar coordinate $(r,\theta)=\left(\sqrt{x^2+y^2},\text{arctan}\frac{y}{x} \right)$.

As widely known, electromagnetic (EM) energy will radiate from an antenna and propagate through free space as a sequence of ever-expanding spherical wavefronts. Therefore, in general, the channel from the
user to the $n$th BS antenna can be modeled as \cite{zhou2015spherical,jiang2005spherical,cui2021nearWB} 
\begin{equation}
[\textbf{h}(x,y)]_{n}=\sqrt{\left( \frac{c}{4\pi f_c r^{(n)}}\right)^2}e^{-j2\pi f_c\frac{r^{(n)}}{c}},
\label{Eq:channel_fs}
\end{equation}
where $f_c$ is the carrier frequency, $c$ is the speed of light, and $r^{(n)}=\sqrt{x^2+(y-(n-1)d)^2}$ denotes the distance between the $n$th antenna and the user. Here we assumed a line of sight (LoS) channel; extension to the multi-path case is straightforward.
The term $\left( \frac{c}{4\pi f_c r^{(n)}}\right)^2$ represents the free-space path loss, which is approximately constant for different antennas. Therefore, denoting the coefficient term as $g$, the channel vector $\textbf{h}(x,y)$ for the ULA can be expressed as
\begin{equation}
\textbf{h}(x,y)=g \left[e^{-j2\pi \frac{f_c }{c}r^{(1)}},e^{-j2\pi \frac{f_c }{c}r^{(2)}},\ldots, e^{-j2\pi \frac{f_c }{c}r^{(N)}}\right]^T.
\label{Eq:Chan_1}
\end{equation}
The channel difference between each antenna element lies in the phase term, where different distances cause different delays at each antenna compared with the reference one. 
By exploiting the correlation of the delays between different antennas, the channel vector in \eqref{Eq:Chan_1} can be further simplified, i.e., the channel may be described by only a few parameters, as described in the following two cases.

\subsection{Far-Field Channel Model}
\label{subsec:FFCM}
\begin{figure}[t]
    \centering
    \includegraphics[width=0.5\textwidth]{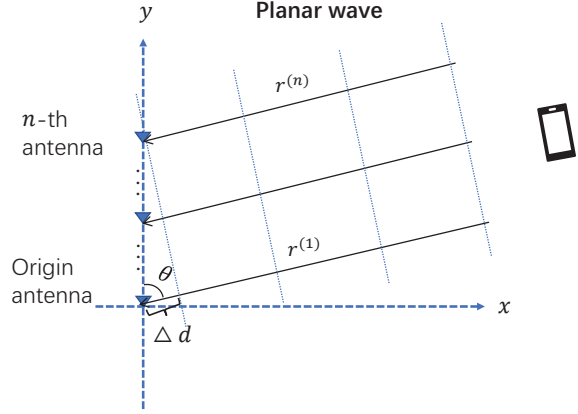}
    \caption{Illustration of the far-field planar-wave model.}
    \label{Planar_wave}
\end{figure}
In classical far-field communications, since the length of the ULA is much smaller than the user distance, the curvature difference of the spherical wavefronts on different antennas is negligible. In this case, as shown in Fig. \ref{Planar_wave}, each antenna element shares the same angle of arrival/departure, and the wavefront is approximately planar.
Therefore, the distance difference from the user to the reference antenna and the $n$th antenna is given by
\begin{equation}
\Delta d=r^{(1)}-r^{(n)}=(n-1)d\text{cos}{\theta}.
\label{Eq:planar_r}
\end{equation}
Then, the $N\times 1$ channel vector $\textbf{h}(x,y)$ in \eqref{Eq:Chan_1} can be rewritten as
\begin{equation}
\textbf{h}(x,y)=g e^{-j2\pi \frac{f_c}{c}r^{(1)}}\textbf{a}(\theta),
\label{Eq:FF_Chan}
\end{equation}
where $\textbf{a}(\theta)$ denotes the array steering vector, given by
\begin{equation}
\textbf{a}(\theta)=\left[ 1, e^{j2\pi \frac{f_c}{c} d \text{cos}\theta}, \cdots, e^{j2\pi \frac{f_c}{c} (N-1)d \text{cos}\theta}\right]^T
=\left[ 1, e^{j\pi \text{cos}\theta}, \cdots, e^{j\pi (N-1)\text{cos}\theta}\right]^T.
\end{equation}

Since $\textbf{h}(x,y)$ is represented by a few parameters, channel estimation is equivalent to the estimation of these parameters. This has been exploited in many works for far-field channel estimation, where parameter estimation is performed based on the theory of compressed sensing (CS) \cite{Yonina_Sampling,Lee2016channel,Ding2018Bayesian,tseng2016enhanced}.
To be more specific,
note that the steering vector $\textbf{a}(\theta)$ is a discrete Fourier vector since 
the phase of its element is linear in the antenna index $n$. Therefore, assuming the angle space is divided into $N$ parts, the channel $\textbf{h}(x,y)$ can be modelled by the following angular-domain sparse model
\begin{equation}
\textbf{h}(x,y)=\textbf{F}\textbf{s},
\label{Eq:FF_Chan2}
\end{equation}
where $\textbf{s}\in \mathcal{C}^{N\times 1}$ is a sparse vector, of which the number of nonzero elements is 1 in the LoS case or $L$ in the $L$ paths case. The matrix
$\textbf{F}=[\textbf{a}(\theta_1), \cdots, \textbf{a}(\theta_N)]\in \mathcal{C}^{N\times N}$ denotes the discrete Fourier transform matrix, where
$\theta_1,\theta_2,\ldots,\theta_N$ denote all the possible angles of arrival of the received signal. Based on the sparse model in \eqref{Eq:FF_Chan2}, various CS algorithms can be applied to estimate $\textbf{s}$. The angle of arrival is then inferred from the nonzero element index of the estimated $\textbf{s}$, and the associated values of these nonzero elements represent the channel coefficients.

\subsection{Near-Field Channel Model}
\label{subsec:NFCM}
\begin{figure}[t]
    \centering
    \includegraphics[width=0.5\textwidth]{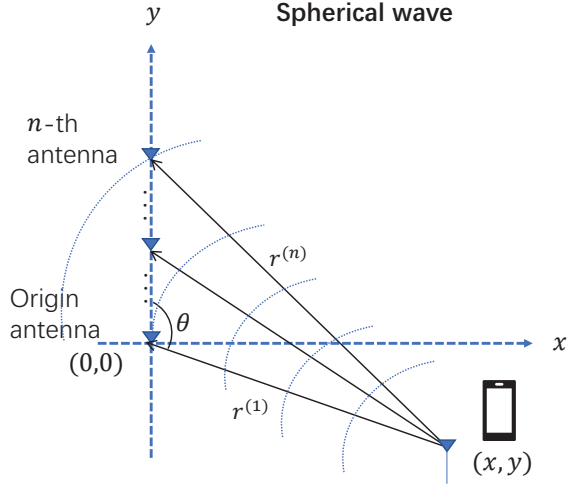}
    \caption{Illustration of the near-field spherical-wave model.}
    \label{Spherical_wave}
\end{figure}

In the radiating near-field region, the aforementioned planar wavefront approximations do not hold. Instead, the wavefront should be accurately modelled under the spherical wavefront assumption \cite{zhang20226g,zhang2022near,cui2022near}, as shown in Fig. \ref{Spherical_wave}.
In this case, the distance between the $n$th antenna and the user is calculated as
\begin{equation}
r^{(n)}=\sqrt{\left(r^{(1)}\right)^2+\left((n-1)d\right)^2-2r^{(1)}(n-1)d\text{cos}\theta}.
\label{Eq:rn}
\end{equation}
Then, the channel vector $\textbf{h}(x,y)$ in \eqref{Eq:Chan_1} is expressed as
\begin{equation}
\textbf{h}(x,y)= g e^{-j2\pi \frac{f_c}{c}r^{(1)}}\textbf{b}_N(\theta,r),
\end{equation}
where the steering vector $\textbf{b}_N(\theta,r)$ is a function of the user angle, $\theta$, and the distance to the reference antenna, $r^{(1)}$. For notational simplicity, we omit the superscript of the distance, and define
\begin{equation}
\textbf{b}_N(\theta,r)=\left[1, e^{-j2\pi \frac{f_c}{c}(r^{(2)}-r^{(1)})},\cdots, e^{-j2\pi \frac{f_c}{c}(r^{(N)}-r^{(1)})}\right]^T.
\label{Eq:vec_b}
\end{equation}
%
Note that the phase of each element in the steering vector $\textbf{b}_N(\theta,r)$ is nonlinear in the antenna index $n$, so that
$\textbf{b}_N(\theta,r)$ cannot be described by a single far-field Fourier vector.

In \cite{cui2022channel}, the authors proposed a polar-domain sparse representation of the near-field channel as
\begin{equation}
\textbf{h}(x,y)=\textbf{D}\textbf{u},
\end{equation}
where $\textbf{u}\in \mathcal{C}^{N\cdot M\times 1}$ is the sparse vector in the polar-domain, with $N$ representing the number of divisions of the angle space $[-1,1]$ and $M$ the number of divisions of the distance range $[r_{\text{min}},r_{\text{max}}]$.
The dictionary $\textbf{D}\in  \mathcal{C}^{N\times N\cdot M}$ is designed by sampling both the angles ($\text{cos}\theta_i\in [-1,1], i=1,\ldots,N$) and distances ($r_i\in [r_{\text{min}},r_{\text{max}}], i=1,\ldots,M$), that is,
\begin{equation}
\textbf{D}=[\textbf{b}_N(\theta_1,r_1), \textbf{b}_N(\theta_1,r_2),\cdots, \textbf{b}_N(\theta_N,r_M)].
\label{Eq:Dic_D}
\end{equation}
The polar-domain sparse representation well describes the spherical wavefront based near-field channel. However,
the employed dictionary $\textbf{D}$ is an angle-distance two-dimensional (2D) dictionary. For such a 2D dictionary, on the one hand, to get finer resolution, the angular and distance sampling interval should be as small as possible, which results in a wide dictionary and thus a high storage burden. On the other hand,
following CS theory \cite{Yonina_Sampling}, to achieve satisfying recovery accuracy, the sampling on angle and distance should make the column coherence of the dictionary as small as possible, which implies a comparatively large sampling interval. Therefore, the dictionary design can be difficult since there is a tradeoff between the dictionary's resolution and coherence. To address this issue, in \cite{cui2022channel}, a nonuniform distance sampling method was proposed to limit the column coherence above a given threshold. However, conceptually, the obtained dictionary is still a wide 2D dictionary, especially when the distance range is large. 
%
%
\section{Near-Field Channel Sparse Representation and Estimation}
\label{sec:CE_Alg}
To address the challenges arising from the polar-domain representation, in this section, we first propose a distance-parameterized angular-domain sparse representation for the near-field channel in Section \ref{subsec:SR_NFC}. Based on this model, joint dictionary learning and sparse recovery based channel estimation algorithms are proposed respectively for the LoS case and multi-path case in Section \ref{subsec:Est_Algo}. Subsection \ref{subsec:RIP_analysis} provides RIP-based recovery conditions, followed by storage and computational complexity analyses in Subsection \ref{subsec:Analysis}.
\subsection{Distance-Parameterized Angular-domain Sparse Representation of Near-Field Channel}
\label{subsec:SR_NFC}
The aforementioned polar-domain representation well describes the characteristics of spherical wavefronts, however, it fails to exploit the fact that: the angle and distance are coupled.
This implies that it is not necessary to construct an extensive dictionary by sampling both angle and distance. Instead, the sparse channel model can be formulated as
\begin{eqnarray}
\textbf{h}(x,y)=\textbf{W}(\textbf{r})\textbf{s},
\label{Eq:h_jointExpre}
\end{eqnarray}
where $\textbf{s}\in \mathcal{C}^{N\times 1}$ is an angular-domain sparse vector as in the far-field case. Here $\textbf{W}(\textbf{r})\in \mathcal{C}^{N\times N}$ is the distance-parameterized dictionary with the argument vector $\textbf{r}=[r_1,r_2,\ldots,r_N]$. Specifically,
each column of $\textbf{W}(\textbf{r})$ is defined by a given angle $\theta_i$ and a distance parameter $r_i$ that remains to be estimated, that is
\begin{equation}
\textbf{W}(\textbf{r})=[\textbf{b}_N(\theta_1,r_1), \textbf{b}_N(\theta_2,r_2),\cdots, \textbf{b}_N(\theta_N,r_N)],
\end{equation}
where $\textbf{b}_N(\theta_i,r_i)$ is defined in \eqref{Eq:vec_b}.

Note that the dictionary $\textbf{W}(\textbf{r})$ is different from the dictionary $\textbf{D}$ in \eqref{Eq:Dic_D} in that each atom of $\textbf{W}(\textbf{r})$ corresponds to a different angle.
This brings benefits in at least two ways. First, the vast storage burden of the polar-domain 2D dictionary is well addressed. This is because the size of $\textbf{W}(\textbf{r})$ only depends on the angular resolution, which is much smaller than the size of $\textbf{D}$ when the distance sampling point $M$ is large.
Second, comparing to the 2D dictionary, the coherence of the dictionary $\textbf{W}(\textbf{r})$ is much smaller as each atom is assigned a different angle. Specifically, assume the angle sampling interval for both dictionaries is set as\footnote{The angle space $\text{cos}\theta\in [-1,1]$ is assumed to be divided into $N$ parts.} $\frac{2}{N}$, and the distance sampling interval for dictionary $\textbf{D}$ is set as $1$ m, we have the following proposition:

\textit{Proposition 1}:
Denote the correlation between two columns $\textbf{b}_N(\theta_p,r_p)$ and $\textbf{b}_N(\theta_q,r_q)$ as
$\epsilon (\theta_p,\theta_q,r_p,r_q)=\left| \frac{1}{N}\textbf{b}_N^H(\theta_p,r_p)\textbf{b}_N(\theta_q,r_q)\right|$,
the coherence of dictionary $\textbf{W}(\textbf{r})$ and dictionary $\textbf{D}$ are respectively given by
\begin{eqnarray}
\epsilon_{\textbf{W}}(\theta_p,\theta_q,r_p)\approx \left|\frac{1}{N} \sum_{n=1}^{N} e^{j2\pi \frac{f_c}{c} \left(
-\frac{4(n-1)d}{N\left(2-(n-1)dz+(n-1)^2d^2z\right)}\right)}
\right|,
\label{Eq:p1_epW}
\end{eqnarray}
and
\begin{eqnarray}
\epsilon_{\textbf{D}}(\theta_p,r_p,r_q)\approx \left|\frac{1}{N} \sum_{n=1}^{N} e^{j2\pi \frac{f_c}{c} \left(
\frac{1}{1+(n-1)^2d^2 x}-1
\right)}
\right|,
\label{Eq:p1_epD}
\end{eqnarray}
where $z=\frac{\text{cos}\theta_p+\text{cos}\theta_q}{r_p^{(1)}}$, and 
$x=\frac{1-\text{cos}^2\theta_p}{2r_p^{(1)}r_q^{(1)}}$.

\textit{Proof:} The proof is provided in Appendix \ref{sec:App_B}.

\begin{figure}[t!]
    {\begin{minipage}[h]{0.47\textwidth}
			\centering
		\centerline{\epsfig{figure=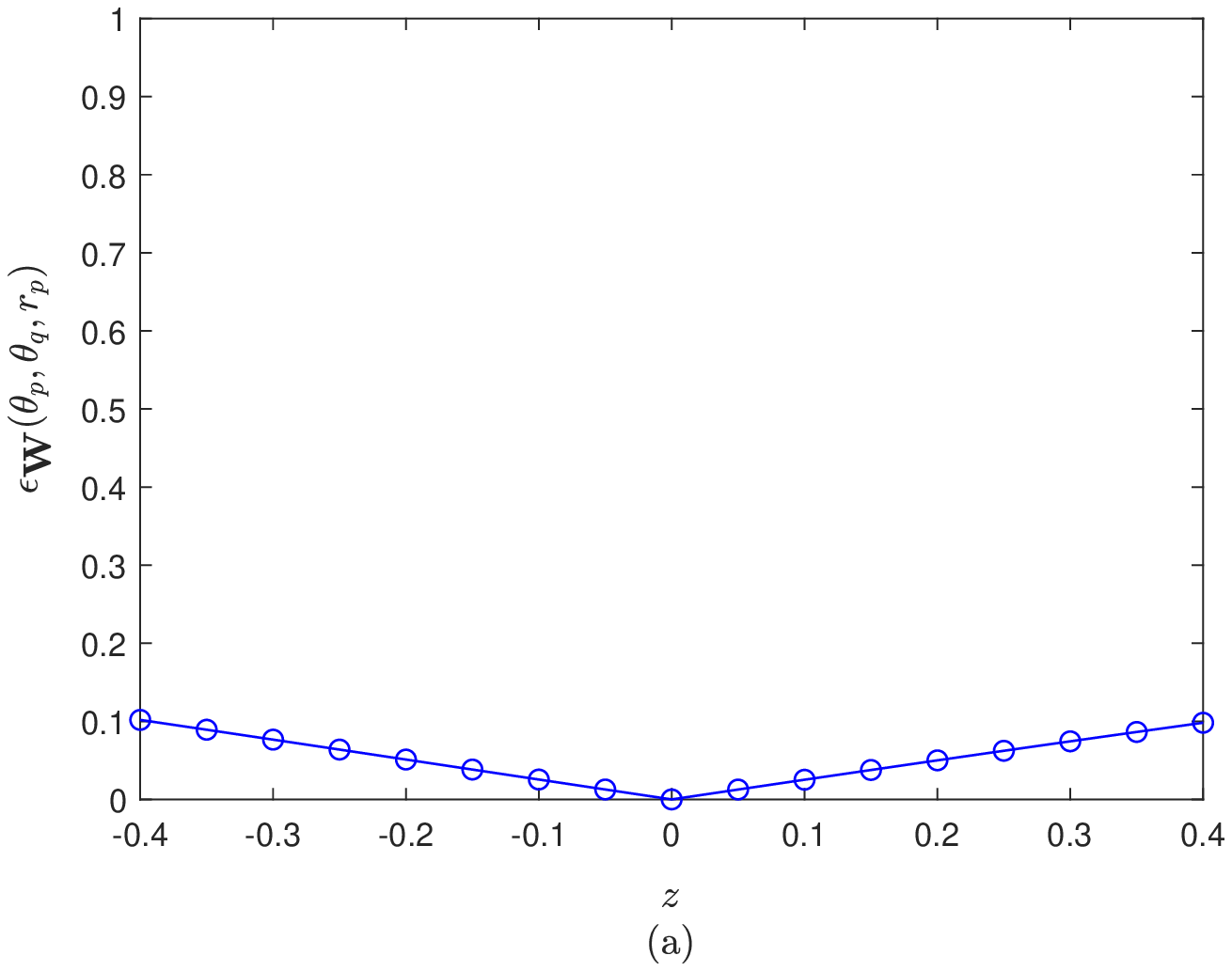,width=1.12\columnwidth}}
		\end{minipage}
	}
    {\begin{minipage}[h]{0.47\textwidth}
			\centering
		\centerline{\epsfig{figure=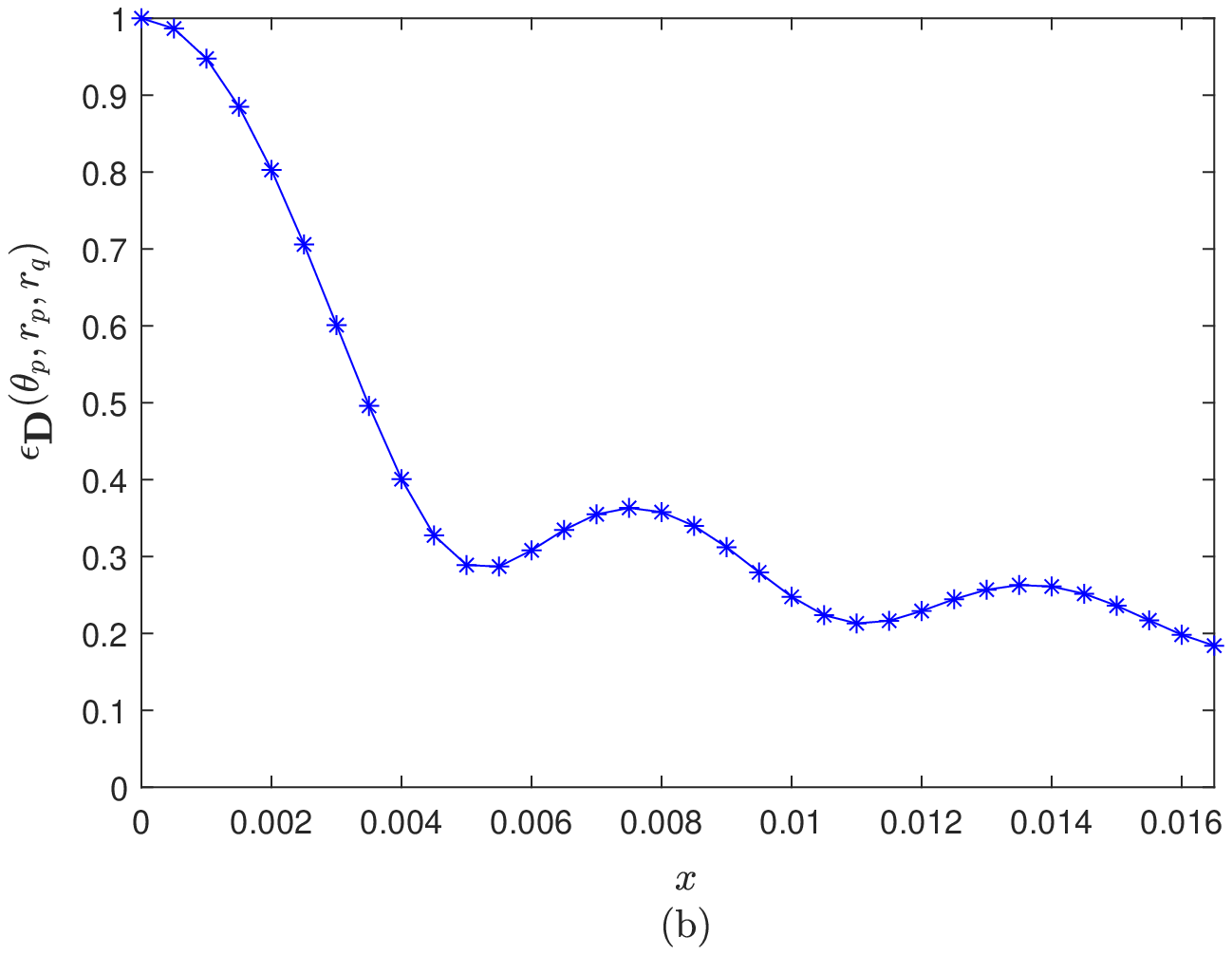,width=1.12\columnwidth}}
		\end{minipage}
	}
	\hfill
    \caption{ Illustration of (a) $\epsilon_{\textbf{W}} (\theta_p,\theta_q,r_q)$ versus $z$, (b) $\epsilon_{\textbf{D}} (\theta_p,r_p,r_q)$ versus $x$.}
\label{Fig_epsilon}
\end{figure}

From the proof, \eqref{Eq:p1_epW} occurs when $r_p^{(1)}=r_q^{(1)}$ and the angle sampling interval is set as  $\text{cos}\theta_p-\text{cos}\theta_q=\frac{2}{N}$. When 
the user distance is comparatively much larger than the antenna array length, i.e., $z\rightarrow 0$, we approximately have
$\epsilon_{\textbf{W}} (\theta_p,\theta_q,r_p)\approx 0$. In this case, columns of $\textbf{W}(\textbf{r})$ are almost orthogonal. While the value of $z$ is not neglectable, the value of $\epsilon_{\textbf{W}}(\theta_p,\theta_q,r_p)$ varies with $z$. Here we illustrate the correlation
numerically in Fig. \ref{Fig_epsilon} (a), where the BS antenna number is set as $N=256$, the carrier frequency is $28$ GHz. In this setting, the theoretical near-field range is in $[3.4558, 351.0857]$ m, and the antenna aperture is $1.3661$ m. We assume the nearest user is located at $5$ m and thus $z$ varies from $-0.4$ to $0.4$. We see that the column correlation is comparatively low over the entire range of $z$. 

\eqref{Eq:p1_epD} occurs when $\theta_p=\theta_q$ and the distance sampling interval is set as $r_p^{(1)}-r_q^{(1)}=1$ m. In the case when $x\rightarrow 0$, we have 
$\epsilon_{\textbf{D}}(\theta_p,r_p,r_q)\approx 1$, indicating a high column coherence of the dictionary $\textbf{D}$. In other cases, with the setting described above, $x$ varies from $0$ to $0.0167$, we plot $\epsilon_{\textbf{D}}(\theta_p,r_p,r_q)$ with respect to $x$ in Fig. \ref{Fig_epsilon} (b). It shows that $\epsilon_{\textbf{D}}(\theta_p,r_p,r_q)$ is non-negligible over the entire range of $x$. 

\begin{figure}[t!]
    {\begin{minipage}[h]{0.47\textwidth}
			\centering
		\centerline{\epsfig{figure=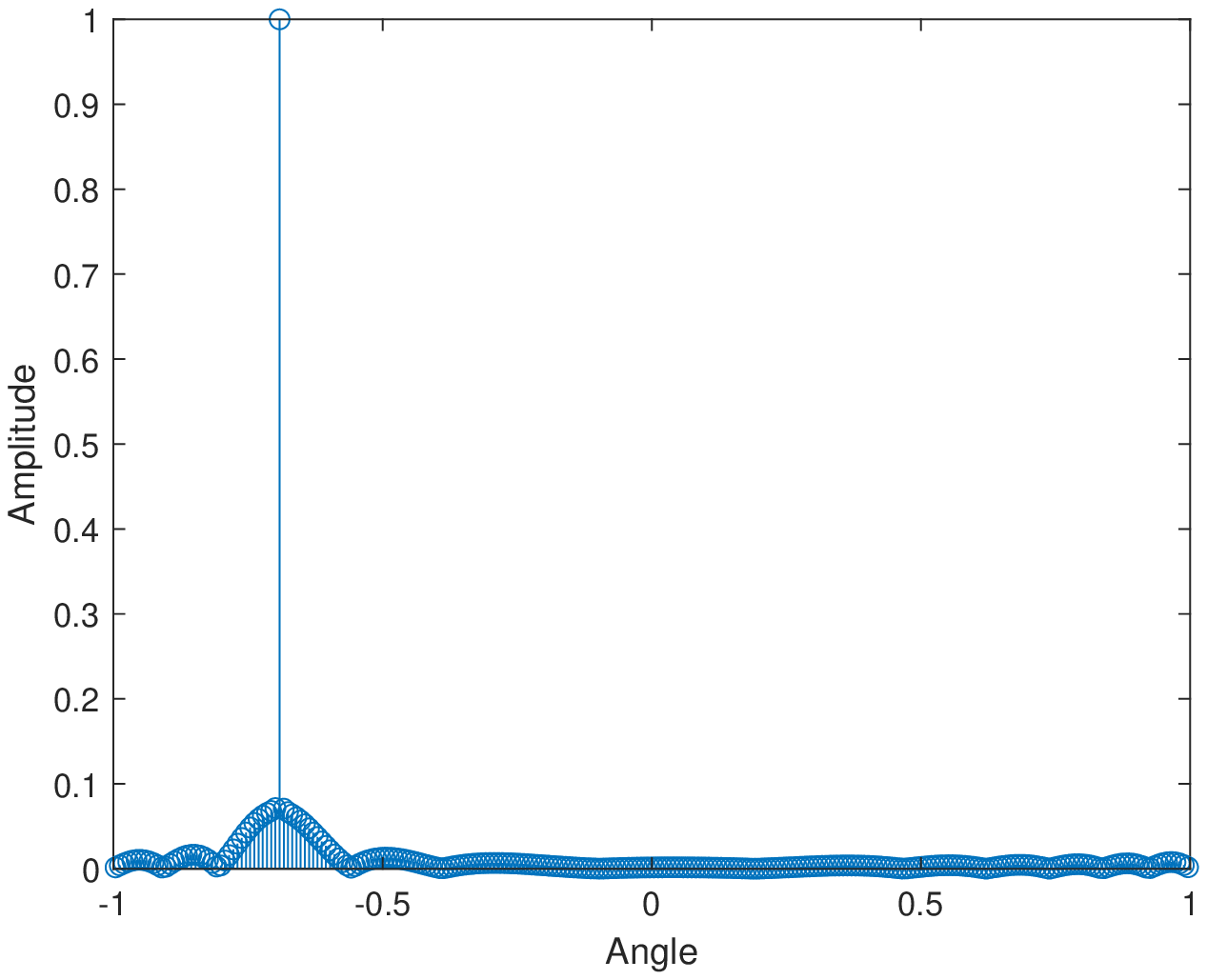,width=1.12\columnwidth}}
		\end{minipage}
	}
    {\begin{minipage}[h]{0.47\textwidth}
			\centering
		\centerline{\epsfig{figure=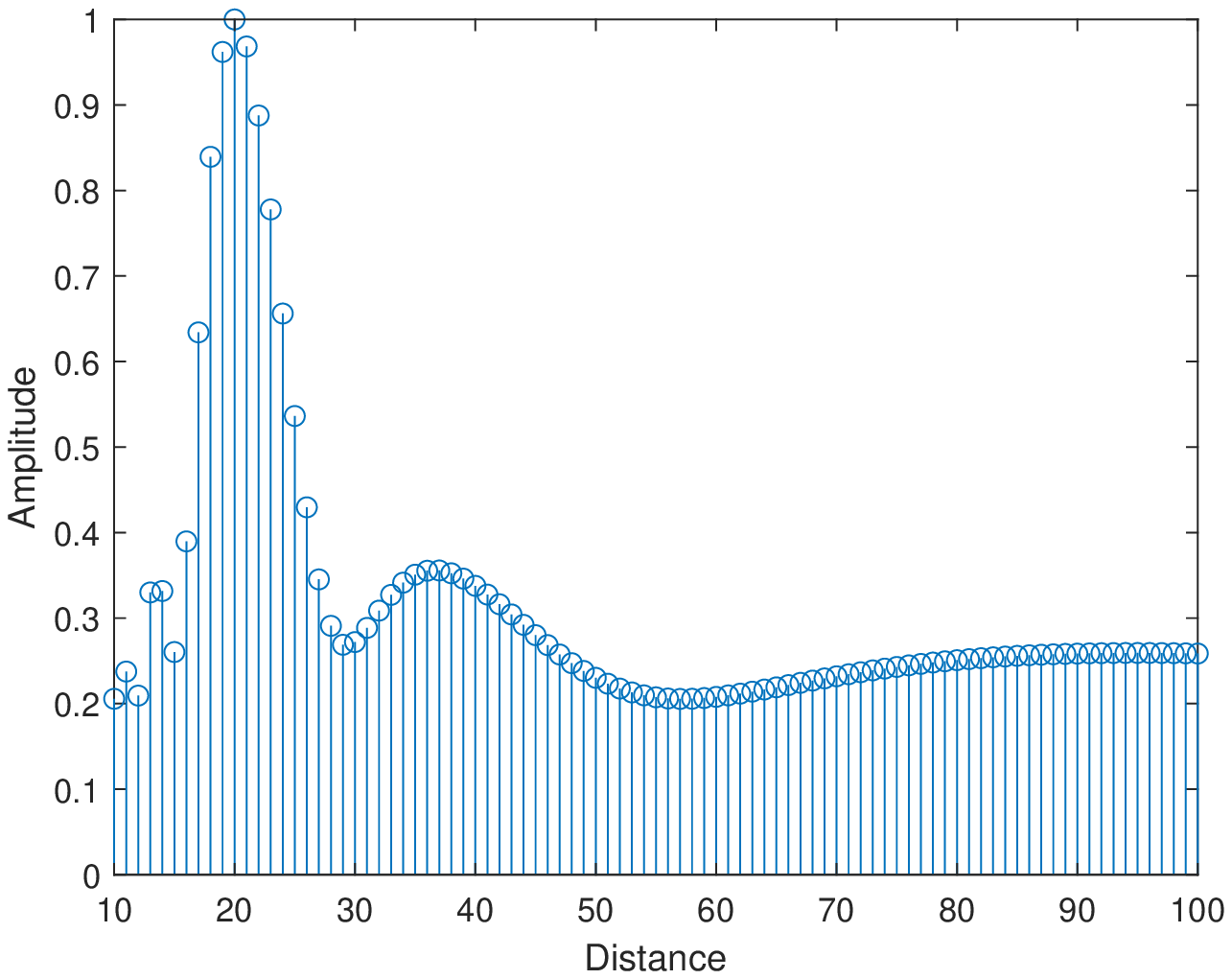,width=1.12\columnwidth}}
		\end{minipage}
	}
	\hfill
    \caption{ Dictionary coherence comparison. (a) The proposed dictionary. (b) The polar-domain dictionary.}
\label{Fig:correlation}
\end{figure}

From the above analysis, it is evident that the proposed dictionary 
$\textbf{W}(\textbf{r})$ has inherent superiority over the 
polar-domain 2D dictionary in the sense of dictionary coherence. This is further supported by Fig. \ref{Fig:correlation} where the simulation setting is the same as that of Fig. \ref{Fig_epsilon}.
In particular,
in Fig. \ref{Fig:correlation} (a), all columns of $\textbf{W}(\textbf{r})$ are parameterized by the same distance.
While Fig. \ref{Fig:correlation} (b) shows the correlation of columns in dictionary $\textbf{D}$ with the same angle but different distances \footnote{For clarity, we show only a portion of the distance}.
We see that for dictionary $\textbf{W}(\textbf{r})$, the coherence is much smaller than that of the 2D dictionary which samples both the angle and distance space, in line with the previous theoretical analysis.

The problem now is to simultaneously estimate the distance $\textbf{r}$ of the dictionary and recover the sparse vector $\textbf{s}$. We consider the following two cases of the near-field channel: LoS and multi-path.
\subsection{Joint Dictionary Learning and Sparse Recovery based Channel Estimation}
\label{subsec:Est_Algo}
\subsubsection{Line of Sight Channel Estimation}
\label{subsec:LoS_CE}
In this case, we only take into account the signal propagating from each user to the BS in a straight line. In the channel acquisition stage, different users transmit mutually orthogonal pilots, so that the channel estimation can be performed individually for each user. 
Without loss of generality, let $p$ be a transmitted pilot from a user located at $(x,y)$, the received signal of the $N$ antennas at the BS is given by
\begin{equation}
\textbf{y}=\textbf{h}(x,y) p+\textbf{v},
\end{equation}
where $\textbf{v}\in \mathcal{C}^{N\times 1} $ is the Gaussian noise of $N$ receive antennas. 
The channel $\textbf{h}(x,y)$ is assumed to be invariant during one or several blocks of transmission time. The pilot  $p$ is known at the receiver and thus can equivalently be assumed as 1 for simplicity \cite{cui2022channel}. In case when each antenna is connected with a radio frequency (RF) chain, i.e., the fully digital receiver, the length of the observation is equivalent to the length of the channel. Therefore, the estimation of $\textbf{h}(x,y)$ can be performed by conventional methods like the least square algorithm. 

\begin{figure}[t]
    \centering
    \includegraphics[width=0.5\textwidth]{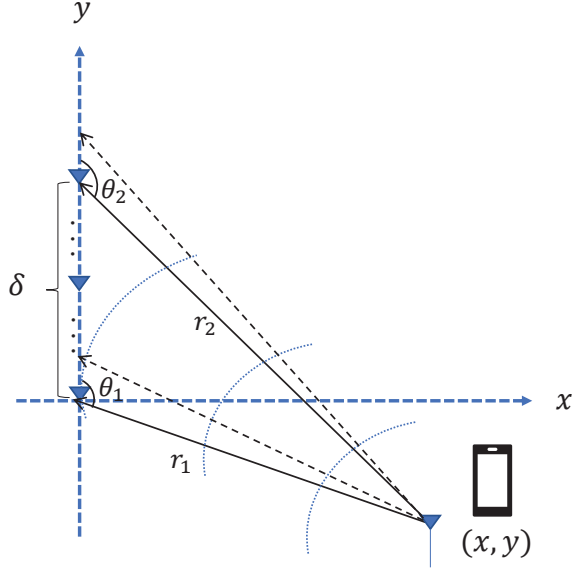}
    \caption{Illustration of the antenna selection for near-field LoS channel estimation.}
    \label{los2G}
\end{figure}

In practice, however, to reduce the cost and complexity of the hardware, only a few number of radio frequency (RF) chains, denoted by $N_{\rm RF}$,  are used to serve massive antennas, i.e., $N_{\rm RF}\ll N$ \cite{molisch2017hybrid,bogale2014beamforming}. In this case, using conventional methods to estimate $\textbf{h}(x,y)$ causes large pilot overhead since it requires $N_{\rm RF}T\ge N$, where $T$ is the pilot length. One solution is to explore the sparsity of the channel, as we analyzed before, and using compressed sensing methods \cite{Yonina_CompressedSensing}.
Here, we represent the channel by the proposed parameterized sparse model as in \eqref{Eq:h_jointExpre}, and adopt 
the simple switching architecture to connect these $N$ antennas and $N_{\rm RF}$ RF chains \cite{Molisch_2004,mendez2015channel,gao2017massive}.
The switches will select $N_{\rm RF}$ antennas during each pilot time, therefore, a total of $P=N_{\rm RF}T$ antennas are selected over time $T$. Generally, we have $P<N$, thus the pilot overhead is reduced. In our method, we select two subset of antennas and each with length $\frac{P}{2}$, as shown in Fig. \ref{los2G}. The distance between the reference antennas of these two subset is denoted as $\delta$. The angles and the distances between the user and the reference antennas of the two subsets are respectively denoted as $\theta_1$, $\theta_2$ and $r_1$, $r_2$. The measurement for each subset is given by 

%
%
\begin{equation}
\textbf{y}_j=\textbf{h}_j(x,y)+\textbf{v}_j=\textbf{W}_j(\textbf{r})\textbf{s}_j+\textbf{v}_j, \quad j=1,2,
\label{Eq:RecSignal}
\end{equation}
where $\textbf{h}_j (x,y)\in \mathcal{C}^{\frac{P}{2}\times 1}$ represents the channel between the user and the $j$th subarray\footnote{Note that each subarrary uses its own reference antenna in representing the channel.}.  
$\textbf{W}_j(\textbf{r})\in \mathcal{C}^{\frac{P}{2}\times N}$ is the corresponding sparse representation dictionary,
and
$\textbf{s}_j\in \mathcal{C}^{N\times 1}$ is the associated angular-domain sparse vector. 

Based on \eqref{Eq:RecSignal}, now the joint dictionary learning and sparse recovery based channel estimation can be performed in the following way. On one hand, 
for given dictionary $\textbf{W}_j(\textbf{r})$, the estimation of $\textbf{s}_j$ can be done by various compressed sensing algorithms such as basis pursuit (BP) \cite{chen1994basis}, iterative soft thresholding algorithm (ISTA)\cite{beck2009fast} and orthogonal matching pursuit (OMP)\cite{tropp2007signal}. Here we adopt the OMP algorithm which is widely used for its simplicity.
On the other hand, as shown in Fig. \ref{los2G}, once we obtained the estimates of $\theta_1$ and $\theta_2$, the unknown distance can be calculated by law of sine, that is
\begin{equation}
\frac{r_1}{\text{sin}\theta_2}=\frac{\delta}{\text{sin}(\theta_2-\theta_1)}=\frac{r_2}{\text{sin}\theta_1}.
\label{Eq:r_cal}
\end{equation}
The estimated distance is then used for the dictionary updating. This process iterates until certain conditions are met, and we summarized it as the dictionary learning (DL)-OMP algorithm, as shown in \textit{Algorithm 1}.
\begin{table}[t!]
\normalsize
\centering
 \label{Alg:LOS}
  \begin{tabularx}{\linewidth}{XXXXXXX}
\toprule
 \textbf{Algorithm 1} The proposed DL-OMP algorithm for near-field LoS channel estimation\\
\midrule
  \textbf{Inputs:}\\
  \vspace{-0.1cm}
   Received signal $\textbf{y}_1$ and $\textbf{y}_2$; the initial dictionary $\textbf{W}^{\circ}_1$ and $\textbf{W}^{\circ}_2$; the maximum and the minimum distances $r_\text{max}$ and $r_\text{min}$; antenna spacing between the two subarrays $\delta$; dictionary update iterations $K_\text{iter}$
   \vspace{0.1cm}
   \\
  \textbf{Output:}\\
  \vspace{-0.1cm}
  The estimated angle $\hat{\theta}$, distance $\hat{r}$ and the near-field channel $\hat{\textbf{h}}(x,y)$.\\
  \vspace{0.1cm}
  1. \textbf{for} $k\in\{1,2,\cdots,K_\text{iter}\}$ \textbf{do}\\
  2. Estimating the angle $\hat{\theta}_1$:  $\hat{\theta}_1\leftarrow \textbf{W}_1^H \textbf{y}_1$.\\
  3. Selecting a sub-dictionary $\textbf{W}_{2,\text{sub}}$ from 
  $\textbf{W}_{2}$ based on \eqref{Eq:sel_W2}.\\
  4. Estimating the angle $\hat{\theta}_2$:  $\hat{\theta}_2\leftarrow \textbf{W}_{2,\text{sub}}^H \textbf{y}_2$.\\
  5. Calculating distances $\hat{r}_1$ and $\hat{r}_2$: $\hat{r}_1=\frac{\delta \text{sin}\hat{\theta}_2}{\text{sin}(\hat{\theta}_2-\hat{\theta}_1)}$, $\hat{r}_2=\frac{\delta \text{sin}\hat{\theta}_1}{\text{sin}(\hat{\theta}_2-\hat{\theta}_1)}$.\\
  6. Calculating the updating vectors $\boldsymbol\alpha_1$, $\boldsymbol\alpha_2$ based on \eqref{Eq:alpha1} and \eqref{Eq:alpha2}.\\
  7. Updating dictionaries $\textbf{W}_1$ and $\textbf{W}_2$ as
   \begin{eqnarray}
    \textbf{W}_1=\text{diag}(\boldsymbol\alpha_1)\textbf{W}^{\circ}_1, \quad \textbf{W}_2=\text{diag}(\boldsymbol\alpha_2)\textbf{W}^{\circ}_2, \nonumber
   \end{eqnarray}
  8. \textbf{end for} \\
  9. Estimating the channel coefficient $\hat{\tilde{g}}$ based on $\hat{\theta}=\hat{\theta}_1$, $\hat{r}=\hat{r}_1$ by \eqref{Eq:g}, and reconstructing the channel vector $\hat{\textbf{h}}(x,y)$ as in \eqref{Eq:channel_h}.
   \\
  10. \textbf{return} $\hat{\theta}$, $\hat{r}$ and $\hat{\textbf{h}}(x,y)$.
  \\
  \bottomrule
 \end{tabularx}
\end{table}
In particular, 
assume the user is located in $[r_{\text{min}},r_{\text{max}}]$ and the angle space is divided into $N$ parts, the initial dictionary $\textbf{W}^{\circ}_1$ and $\textbf{W}^{\circ}_2$ are both constructed with $r=r_{\text{max}}$ for each column $\textbf{b}_P(\theta_i,r_{\text{max}})$, $i=1,2,\ldots, N$, i.e.,
\begin{equation}
\textbf{b}_P(\theta_i,r_{\text{max}})=\left[1,e^{-j2\pi \frac{f_c}{c}(r_{\text{max}}^{(2)}-r_{\text{max}})},\cdots, e^{-j2\pi \frac{f_c}{c}(r_{\text{max}}^{(P/2)}-r_{\text{max}})}\right]^T,
\label{Eq:vec_b2}
\end{equation}
where $r_{\text{max}}^{(p)}, p=1,\ldots,\frac{P}{2}$ is defined by \eqref{Eq:rn}.

In step 2, the estimation of $\theta_1$, denoted as $\hat{\theta}_1$, is obtained based on the OMP algorithm, that is, calculating the inner products between the received signal and each column of the dictionary to yield $\boldsymbol{\phi}_1=\textbf{W}_1^H \textbf{y}_1$, then obtain $\hat{\theta}_1$ by detecting the position of the maximum value in $\boldsymbol{\phi}_1$.

Steps 3 and 4 are for the estimation of $\theta_2$. In principle, based on $\textbf{y}_2$ and $\textbf{W}_2$, the estimate $\hat{\theta}_2$ can be obtained in a similar way as $\hat{\theta}_1$ in step 2. Here, however, we further reduce the computational complexity by exploiting the information brought by $\hat{\theta}_1$. Specifically, based on $\hat{\theta}_1$, and the user range $[r_{\text{min}},r_{\text{max}}]$, it is possible to get a rough estimation of the range of $\theta_2$, that is
%
\begin{eqnarray}
\text{cos}(\theta_{\text{min}})=\frac{-(d_1^2+\delta^2-r_{\text{min}}^2)}{2d_1\delta},
\nonumber\\
\text{cos}(\theta_{\text{max}})=\frac{-(d_2^2+\delta^2-r_{\text{max}}^2)}{2d_2\delta},
\label{Eq:sel_W2}
\end{eqnarray}
where
\begin{eqnarray}
d_1= \sqrt{ r_{\text{min}}^2 + \delta^2-2r_{\text{min}} \delta \text{cos}\hat{\theta}_1},
\nonumber\\
d_2= \sqrt{ r_{\text{max}}^2 + \delta^2-2r_{\text{max}} \delta \text{cos}\hat{\theta}_1}.\nonumber
\end{eqnarray}
Then, we can select a sub-dictionary from  $\textbf{W}_2$ within this range, denoted as $\textbf{W}_{2,\text{sub}}$, and use it to estimate $\theta_2$ by detecting the maximum elements of 
$\boldsymbol{\phi}_2=\textbf{W}_{2,\text{sub}}^H \textbf{y}_2$,
as in step 4.

In step 5, we calculated the distance estimate $\hat{r}_1$ and $\hat{r}_2$ based on \eqref{Eq:r_cal}.

In steps 6 and 7, we update dictionaries $\textbf{W}_1$ and $\textbf{W}_2$ as in the following proposition:

\textit{Proposition 2:} 
With distance estimates $\hat{r}_1$ and $\hat{r}_2$, dictionaries $\textbf{W}_1$ and $\textbf{W}_2$ can be updated as
\begin{equation}
\textbf{W}_1=\text{diag}(\boldsymbol\alpha_1)\textbf{W}^{\circ}_1,
\end{equation}
\begin{equation}
\textbf{W}_2=\text{diag}(\boldsymbol\alpha_2)\textbf{W}^{\circ}_2,
\end{equation}
where
\begin{equation}
\boldsymbol\alpha_1=\left[1, e^{-j2\pi \frac{f_c}{c}(d^2(1-\text{cos}^2\hat{\theta}_1))(\frac{1}{2\hat{r}_1}-\frac{1}{2r_{\text{max}}})},\cdots, e^{-j2\pi \frac{f_c}{c}((P/2-1)^2d^2(1-\text{cos}^2\hat{\theta}_1))(\frac{1}{2\hat{r}_1}-\frac{1}{2r_{\text{max}}})}\right]^T.
\label{Eq:alpha1}
\end{equation}
\begin{equation}
\boldsymbol\alpha_2=\left[1, e^{-j2\pi \frac{f_c}{c}(d^2(1-\text{cos}^2\hat{\theta}_2))(\frac{1}{2\hat{r}_2}-\frac{1}{2r_{\text{max}}})},\cdots, e^{-j2\pi \frac{f_c}{c}((P/2-1)^2d^2(1-\text{cos}^2\hat{\theta}_2))(\frac{1}{2\hat{r}_2}-\frac{1}{2r_{\text{max}}})}\right]^T.
\label{Eq:alpha2}
\end{equation}
%

\textit{Proof:} The proof is provided in Appendix \ref{App_C}.
%

In step 9, with the angle and distance estimates $\hat{\theta}=\hat{\theta}_1$ and $\hat{r}=\hat{r}_1$, the channel coefficient, $g$, can be estimated by solving
\begin{equation}
\hat{g}=\min\limits_{g}||\textbf{y}_1-g\tilde{\textbf{b}}_{P}(\hat{\theta},\hat{r})||^2 \quad \rightarrow \quad \hat{g}=\tilde{\textbf{b}}^{\dag}_{P}(\hat{\theta},\hat{r})\textbf{y}_1,
\label{Eq:g}
\end{equation}
where
\begin{equation}
\tilde{\textbf{b}}_{P}(\hat{\theta},\hat{r})=e^{-j2\pi \frac{f_c}{c}\hat{r}}\left[1,e^{-j2\pi \frac{f_c}{c}(\hat{r}^{(2)}-\hat{r})},\cdots,e^{-j2\pi \frac{f_c}{c}(\hat{r}^{(P/2)}-\hat{r})}\right]^T.
\label{Eq:b2P}
\end{equation}
Then, the channel vector can be estimated as
\begin{equation}
\hat{\textbf{h}}(x,y)=\hat{g}e^{-j2\pi \frac{f_c}{c}\hat{r}}\textbf{b}_N(\hat{\theta},\hat{r}).
\label{Eq:channel_h}
\end{equation}
\subsubsection{Multi-path Channel Estimation}
\label{subsec:MP_CE}
In this case, except the LoS path, signals from some scatters are also taken into account. The channel is modelled by a multi-path model as
\begin{equation}
\textbf{h}(x,y)=\sum_{l=1}^{L}g_l e^{-j2\pi \frac{f_c}{c} r_l^{(1)}}\textbf{b}_N(\theta_l, r_l),
\label{Eq:multi-pathC}
\end{equation}
where $g_l$, $\theta_l$ and $r_l$ are respectively the gain, angle of arrival and distance of the $l$th path.
Similar to the LoS case in \eqref{Eq:RecSignal}, the received signal from multi-path channel can also be formulated by a sparse model
\begin{equation}
\textbf{y}_j=\textbf{W}_j(\textbf{r})\textbf{s}_j+\textbf{v}_j, \quad j=1,2.
\end{equation}
The difference is that for the multi-path case, more than one element of $\textbf{s}_j$ are nonzero. This poses new challenges for channel estimation as interference from other paths exists. Multi-path interference reduces the angular estimation accuracy at initial iterations.
To solve this problem,
we propose to update each column of the dictionary separately within each iteration, that is,
\begin{equation}
\textbf{W}_j(:,i)=\text{diag}(\boldsymbol\alpha_{j}(i))\textbf{W}_j^{\circ}(:,i),\quad i=1,2,\ldots,N; j=1,2,
\label{Eq: uptW}
\end{equation}
with
\begin{equation}
\boldsymbol\alpha_{j}(i)=\left[1, e^{-j2\pi \frac{f_c}{c}(d^2(1-\text{cos}^2\theta_i))(\frac{1}{2\hat{r}_{j}}-\frac{1}{2r_{\text{max}}})},\cdots, e^{-j2\pi \frac{f_c}{c}((P/2-1)^2d^2(1-\text{cos}^2\theta_i))(\frac{1}{2\hat{r}_{j}}-\frac{1}{2r_{\text{max}}})}\right]^T.
\label{Eq:alphai}
\end{equation}
details are summarized in \textit{Algorithm 2}.
\begin{table}[t!]
\normalsize
\centering
 \label{Alg:LOS}
  \begin{tabularx}{\linewidth}{XXXXXXX}
\toprule
 \textbf{Algorithm 2} The proposed DL-OMP algorithm for near-field multi-path channel estimation\\
\midrule
  \textbf{Inputs:}\\
  \vspace{-0.1cm}
   Received signal $\textbf{y}_1$ and $\textbf{y}_2$; the initial dictionary $\textbf{W}^{\circ}_1$ and $\textbf{W}^{\circ}_2$; the maximum and the minimum distances $r_\text{max}$ and $r_\text{min}$; antenna spacing between the two subsets $\delta$; dictionary update iterations $K_\text{iter}$; the number of paths $L$
   \vspace{0.1cm}
   \\
  \textbf{Output:}\\
  \vspace{-0.1cm}
  The estimated angles $\Lambda_{\theta}$, distances $\Lambda_{r}$ and the near-field channel $\hat{\textbf{h}}(x,y)$.\\
  \vspace{0.1cm}
  1. Initialize: $\Lambda_{\theta}=\varnothing$, $\Lambda_{r}=\varnothing$,
                 $\textbf{z}_1=\textbf{y}_1$, $\textbf{z}_2=\textbf{y}_2$.\\
  2. \textbf{for} $l\in\{1,2,\cdots,L\}$ \textbf{do}\\
  3. \quad \textbf{for} $k\in\{1,2,\cdots,K_\text{iter}\}$ \textbf{do}\\
  4. \quad Estimating the angle $\hat{\theta}_1$:  $\hat{\theta}_1\leftarrow \textbf{W}_1^H \textbf{z}_1$.\\
  5. \quad Selecting a sub-dictionary $\textbf{W}_{2,\text{sub}}$ from $\textbf{W}_2$ based on \eqref{Eq:sel_W2}.\\
  6. \quad Estimating the angle $\hat{\theta}_2$:  $\hat{\theta}_{2}\leftarrow \textbf{W}_{2,\text{sub}}^H \textbf{z}_2$.\\
  7. \quad Calculating distances $\hat{r}_1$ and $\hat{r}_2$: $\hat{r}_1=\frac{\delta \text{sin}\hat{\theta}_2}{\text{sin}(\hat{\theta}_2-\hat{\theta}_1)}$, $\hat{r}_2=\frac{\delta \text{sin}\hat{\theta}_1}{\text{sin}(\hat{\theta}_2-\hat{\theta}_1)}$.\\
  8. \quad Updating each column of $\textbf{W}_1$ and $\textbf{W}_2$ based on \eqref{Eq: uptW} and \eqref{Eq:alphai}.\\
  9. \quad \textbf{end for} \\
  10. \quad $\Lambda_{\theta}\leftarrow \Lambda_{\theta} \cup \hat{\theta}_1$, $\Lambda_{r}\leftarrow \Lambda_{r} \cup \hat{r}_1$.\\
  11. \quad Constructing the support $\textbf{B}$ based on $\Lambda_{\theta}$ and $\Lambda_{r}$ and calculating the coefficients as in \eqref{Eq:mpcoeff}.\\
  12. \quad Calculating the residual signal: $\textbf{z}_1\leftarrow \textbf{y}_1-\textbf{B}_1\hat{{\textbf{g}}}$,\quad
     $\textbf{z}_2\leftarrow \textbf{y}_2-\textbf{B}_2\hat{{\textbf{g}}}$, where $\textbf{B}_1$ and $\textbf{B}_2$ are \\
     \qquad  subsets of $\textbf{B}$ selected corresponding to the received antenna indices. \\
  13. \textbf{end for} \\
  14. Reconstructing the channel vector $\hat{\textbf{h}}(x,y)$ based on \eqref{Eq:channel_h2}.
   \\
  15. \textbf{return} $\Lambda_{\theta}$, $\Lambda_{r}$ and $\hat{\textbf{h}}(x,y)$.
  \\
  \bottomrule
 \end{tabularx}
\end{table}

After the estimation of each path, the support set $\textbf{B}$ is obtained with its $l$th column given by
$e^{-j2\pi \frac{f_c}{c}\hat{r}_l}\textbf{b}_{N}(\hat{\theta}_l,\hat{r}_l)$.
The coefficient vector $\hat{{\textbf{g}}}=[\hat{g}_1,\ldots,\hat{g}_l]$ can be obtained by solving
\begin{equation}
\hat{{\textbf{g}}}=\min\limits_{{\textbf{g}}}||\textbf{y}_1-\textbf{B}_1{\textbf{g}}||^2 \quad \rightarrow \quad \hat{{\textbf{g}}}=\textbf{B}_1^{\dag}\textbf{y}_1,
\label{Eq:mpcoeff}
\end{equation}
where $\textbf{B}_1$ is a subset of $\textbf{B}$, selected corresponding to the antenna indices of $\textbf{y}_1$.

Then, the multi-path channel can be reconstructed as
\begin{equation}
\hat{\textbf{h}}(x,y)=\sum_{l=1}^{L}\hat{g}_l e^{-j2\pi \frac{f_c}{c}\hat{r}_l}\textbf{b}_{N}(\hat{\theta}_l,\hat{r}_l).
\label{Eq:channel_h2}
\end{equation}
\subsection{The RIP-based Recovery Condition Analysis}
\label{subsec:RIP_analysis}
In general, a dictionary $\boldsymbol{\Phi}$ satisfies the restricted isometry property (RIP) of order $L$ if there exists an isometry constant $\delta_L\in(0,1)$ such that
\begin{equation}
(1-\delta_L)||\textbf{x}||_2^2 \le ||\boldsymbol{\Phi}\textbf{x}||_2^2\le (1+\delta_L)||\textbf{x}||_2^2,
\end{equation}
holds for any $L$-sparse vector $\textbf{x}$ \cite{Davenport_TIT2010}. In other words, $\boldsymbol{\Phi}$ acts as an approximate isometry on the set of vectors that are $L$-sparse.
The RIP for a dictionary provides a sufficient condition to guarantee successful sparse recovery using a wide variety of algorithms, e.g.,
it is shown in \cite{Mo_TIT2012} that the RIP of order $L+1$ (with isometry constant $\delta_{L+1}<\frac{1}{1+\sqrt{L}}$) can be a sufficient condition to permit the OMP algorithm to exactly recover any $L$-sparse signals. 

For the proposed DL-OMP algorithms, as shown in \textit{Algorithm 1} and \textit{Algorithm 2}, the task of each sub-iteration for the $l$th path estimation can be separated into two parts: the angle estimation and the updating of the dictionary. As these two parts affect each other, it is interesting to investigate their dependence. To this end, we provide an analysis for the effect of the distance-induced dictionary perturbation on the angle recovery.
First, define the difference between the dictionary with estimated distance, $\textbf{W}(\hat{\textbf{r}})$, and the dictionary with the true distance, $\textbf{W}(\textbf{r})$, as
\begin{equation}
\Delta \textbf{W}=\textbf{W}(\hat{\textbf{r}})-\textbf{W}(\textbf{r}).
\end{equation}
Based on \eqref{Eq:RecSignal}, we have (where the subscripts are omitted for simplicity)
\begin{equation}
\textbf{y}=(\textbf{W}(\hat{\textbf{r}})-\Delta \textbf{W})\textbf{s}+\textbf{v}=\textbf{W}(\hat{\textbf{r}})\textbf{s}+
\textbf{e},
\end{equation}
where $\textbf{e}=-(\Delta \textbf{W})\textbf{s}+\textbf{v}$. Let $\mu=\frac{||\textbf{e}||_2^2}{||\textbf{s}||_2^2}$, the condition for exact angle recovery is summarized in the following theorem.

\textit{Theorem 1}: If $\textbf{W}(\hat{\textbf{r}})$ satisfies the RIP of order $L+1$ with isometry constant
\begin{eqnarray}
\delta_{L+1}<\frac{1}{2}(\mu+\frac{2}{\sqrt{L}+1})-\sqrt{\frac{1}{4}\mu^2+(\frac{1}{\sqrt{L}+1}+1)\mu },
\label{Eq:bound}
\end{eqnarray}
then the angle of arrival can be exactly recovered by the proposed DL-OMP algorithm.

\textit{Proof:} The proof is provided in Appendix \ref{sec:App_A}.

Discussions:
Note that the variable $\mu$ has a natural interpretation as a ratio between the interference and the signal, where the interference comes from both the distance estimation error and the observation noise.
It can be seen that the term on the right hand side of \eqref{Eq:bound} is a monotonically decreasing function of $\mu$. Therefore, for the proposed DL-OMP,
with the decrease of the distance estimation error, the bound on $\delta_{L+1}$ increases. This implies a more relaxed restricted isometry condition on the dictionary, facilitating the angle recovery in the next iteration.
In the extreme case when $\mu=0$, \eqref{Eq:bound} becomes $\delta_{L+1}<\frac{1}{1+\sqrt{L}}$, which is exactly the recovery condition of the typical OMP algorithm, as analyzed in \cite{Mo_TIT2012}.

Furthermore, in the LoS case, we have $L=1$, \eqref{Eq:bound} is then simplified to
\begin{eqnarray}
\delta_{2}<\frac{1}{2}(\mu+1)-\sqrt{\frac{1}{4}\mu^2+\frac{3}{2}\mu }.
\end{eqnarray}
Since $\delta_2\in (0,\frac{1}{2})$ by definition, we have $0<\mu<0.25$, that is, the maximum allowed channel noise and distance estimation error induced interference is around one-quarter of the energy of the sparse signal.
%
%
%
%
\subsection{Storage and Computational Complexity Analyses}
\label{subsec:Analysis}
The storage space requirement and computational complexity of the proposed DL-OMP algorithm are analyzed and compared with two polar-domain OMP algorithms.  One is the uniform P-OMP algorithm which uniformly samples across the entire angle and distance space.  The other is the nonuniform P-OMP algorithm proposed by \cite{cui2022channel}, which uniformly samples across the angle space but samples the distance nonuniformly.
For a fair comparison, we consider the following parameter settings.
The number of BS antennas is $N$, users are located within $[r_{\text{min}},r_{\text{max}}]$, and $P=TN_{\text{RF}}$ pilots are used for the channel estimation. 
In such a setting, for all these three algorithms, the angle space ($\text{cos}\theta\in[-1,1]$) is divided into $N$ parts. While the distance range $[r_{\text{min}},r_{\text{max}}]$ is sampled with $M$ samples for uniform P-OMP and with $\tilde{M}$ samples for nonuniform P-OMP. Then, it is straightforward to analyze the storage overhead and the computational complexity for each algorithm, as summarized in Table \ref{Tab:I} and detailed in the following. 

The storage requirement is determined by the size of the employed dictionary, we see that the proposed DL-OMP algorithm is conceptually different from the other two polar-domain algorithms, i.e., the proposed dictionary $\textbf{W}(\textbf{r})$ only samples on the angular-domain. Thus the storage space is much smaller compared with the two polar-domain 2D dictionaries, especially when the near-field range is large.

While for the computational complexity, the overall complexity of the proposed DL-OMP algorithm is mainly determined by estimating the angles and updating the dictionary, i.e., steps $2, 4, 7$ in \textbf{Algorithm 1} or steps $4, 6, 8$ in \textbf{Algorithm 2}. Thus we only consider these steps and take the dictionary $\textbf{W}_1$ as an example. Since $\textbf{W}_1\in \mathcal{C}^{\frac{P}{2}\times N}$, the computational complexity for estimating $\hat{\theta}_1$ is $\mathcal{O}(\frac{1}{2}PN)$.
%
%
For the dictionary updating, in the LoS case, the same updating vector is used for each column of $\textbf{W}_1$, the computational complexity is $\mathcal{O}(\frac{1}{2}PN)$ for updating the dictionary.
In the multi-path case, for each column, a $\frac{1}{2}P\times1$ updating vector is calculated and then utilized for dictionary updating, the computational complexity is $\mathcal{O}(PN)$.
Therefore, the overall complexity is approximately $\mathcal{O}(\frac{3}{2}PN)$ in one iteration.
We assume the same computations are required for the estimation of $\theta_2$ and the corresponding dictionary updating, though only a sub-dictionary is selected in the algorithm.  
Then, for $L$ paths and each with $K_{\text{iter}}$ iterations, the overall computational complexity is approximately $\mathcal{O}(3LK_{\text{iter}}PN)$. From experiments, $K_{\text{iter}}=3$ is enough for an accurate channel estimate. The path number $L$ is also small for near-field channels \cite{zhang2022near}. Therefore, the complexity is mainly determined by the pilot length $P$, and the number of antennas $N$, that is, the angular resolution.

For the uniform P-OMP method, its complexity is determined by searching a wide 2D dictionary, of which the number of columns is $NM$. The computational complexity is approximately $\mathcal{O}(LPNM)$. While for the nonuniform P-OMP method, as analyzed in \cite{cui2022channel}, the overall complexity is approximately $\mathcal{O}(LPN\tilde{M})$.
For near-field channel estimation, especially when a large number of antennas and high carrier frequency are employed, we usually have $3K_{\text{iter}}<\tilde{M}<M$. Therefore, the proposed DL-OMP also benefits from lower computational complexity compared with the other two algorithms.
\begin{table}[t!]
	\centering
	\caption{Storage and computational complexity comparison.}
	\label{Tab:I}
\setlength{\tabcolsep}{1.5 mm}{
		\begin{tabular}{|c|c|c|}
	    \hline
         \rule{0pt}{12pt}  Algorithm & Dictionary size & Computational complexity $\mathcal{O}(\cdot)$  \\
        \hline
        \rule{0pt}{12pt} DL-OMP & $\textbf{W}(\textbf{r})\in \mathcal{C}^{N\times N}$ & $\mathcal{O}(3LK_{\text{iter}}PN)$\\
        \hline
        \rule{0pt}{12pt} Uniform P-OMP & $\textbf{D}\in \mathcal{C}^{N\times NM}$ & $\mathcal{O}(LPNM)$ \\
        \hline
        \rule{0pt}{12pt} Nonuniform P-OMP & $\textbf{D}\in \mathcal{C}^{N\times N\tilde{M}}$ & $\mathcal{O}(LPN\tilde{M})$ \\
       \hline
  \end{tabular} }
\end{table}
%

%
%
%
%
\section{Simulation Results}
\label{sec:Sim}
In order to evaluate the proposed algorithms for near-field channel estimation, numerical experiments were conducted in a multi-user orthogonal frequency division multiplexing (OFDM) communication system with default configurations given by Table \ref{Tab:II}.
\begin{table}[t!]
	\centering
	\caption{Simulation Configurations}
	\label{Tab:II}
\setlength{\tabcolsep}{1.5 mm}{
		\begin{tabular}{|c|c|}
	    \hline
         \rule{0pt}{12pt}   The number of BS antennas $N$ & 256\\
        \hline
        \rule{0pt}{12pt} Carrier frequency $f_c$ & 28 GHz\\
        \hline
        \rule{0pt}{12pt} Bandwidth  & 100 MHz\\
        \hline
        \rule{0pt}{12pt} The number of subcarriers & 64 \\
        \hline
        \rule{0pt}{12pt} The number of users & 4 \\
         \hline
        \rule{0pt}{12pt} The number of iterations $K_{\text{iter}}$ & 3\\
         \hline
        \rule{0pt}{12pt} The number of pilots $P$ & 128\\
         \hline
        \rule{0pt}{12pt}
        The angle space of the user & $(-1,1)$\\
         \hline
        \rule{0pt}{12pt} The distance range of the user & $[5,70]$m\\
       \hline
  \end{tabular} }
\end{table}
The performance was evaluated by the normalized mean square error (NMSE) of the channel, that is, $E\left[\frac{||\textbf{h}-\hat{\textbf{h}}||^2_2}{||\textbf{h}||^2_2}\right]$, as shown in the following figures. All the simulation results were obtained by averaging over 5,000 independent trials.

\begin{figure}[t]
    \centering
    \includegraphics[width=0.6\textwidth]{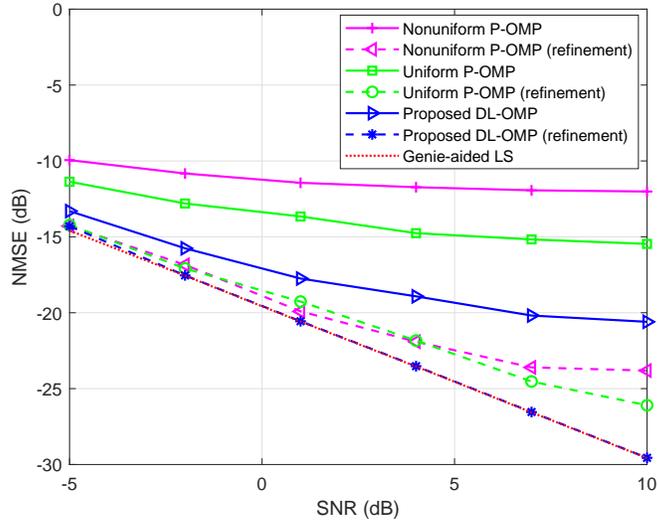}
    \caption{The NMSE of the channel estimate against different SNRs.}
    \label{Fig:Sim_1}
\end{figure}

\begin{figure}[t]
    \centering
    \includegraphics[width=0.6\textwidth]{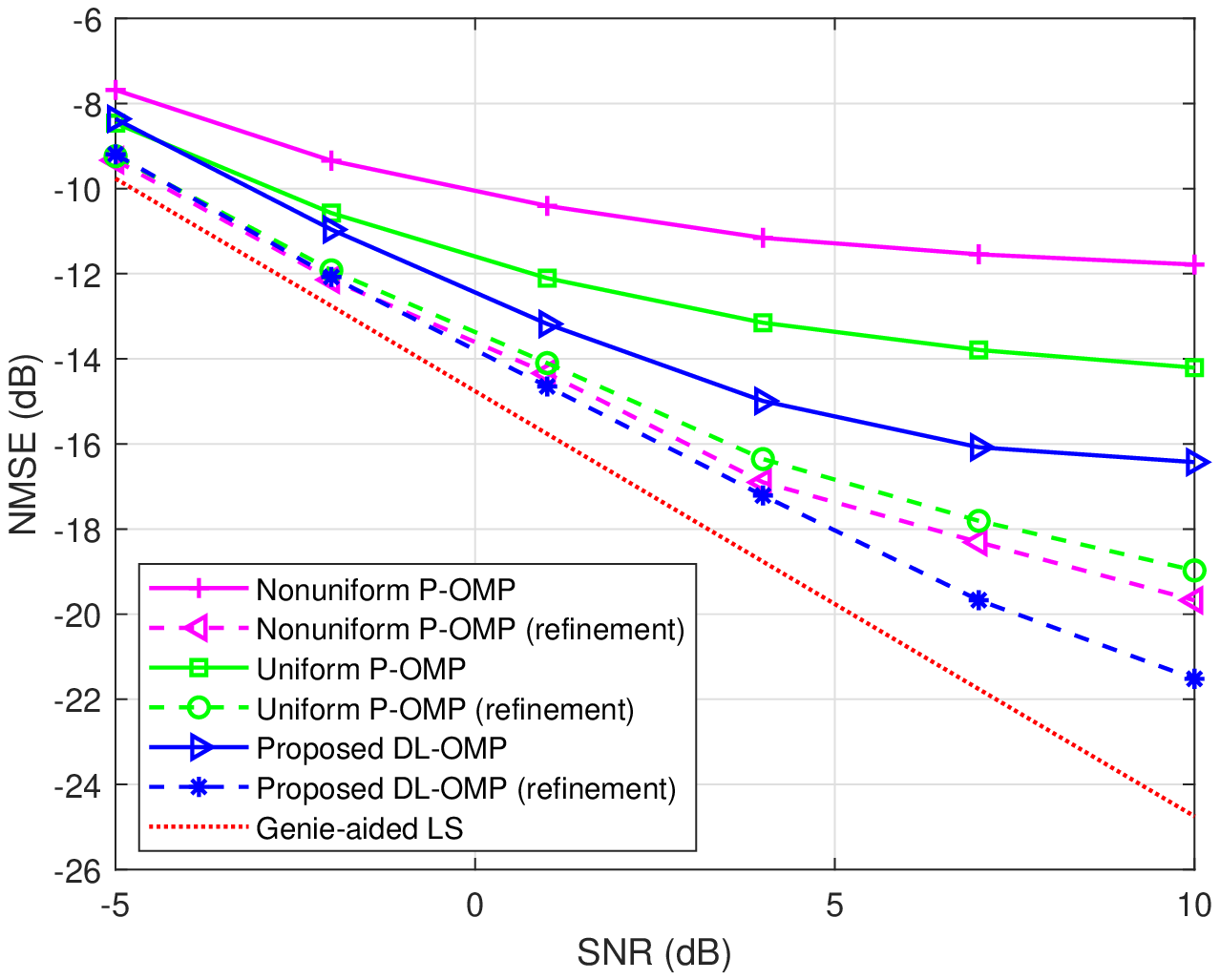}
    \caption{The NMSE of the channel estimate against different SNRs ($L=3$).}
    \label{Fig:Sim_2}
\end{figure}

Fig. \ref{Fig:Sim_1} and Fig. \ref{Fig:Sim_2} respectively illustrate the obtained results of the proposed DL-OMP algorithm with respect to SNRs for the LoS and multi-path channels.
The radiating near-field channel is basically dominated by LoS models. In multi-path settings, the received signals from scatters are also minimal \cite{zhang2022near}, therefore, we set  $L=3$ in Fig. \ref{Fig:Sim_2}. 
Both uniform and nonuniform P-OMP algorithms were plotted to provide comparisons, and the distance sampling interval of the uniform dictionary was set as $2$ m. The genie-aided least square (LS) algorithm, which assumes known angle and distance, served as a benchmark.
We provided two sets of results: initial results by the above algorithms and the refined ones. The refinement was performed by alternatively optimizing the algorithm-obtained angles and distances to minimize the error of the reconstructed signal,
as detailed in \cite{cui2022channel}. 
From these figures, we see that the proposed DL-OMP algorithm outperforms the uniform and nonuniform P-OMP algorithms in the channel estimation, evidenced by superior NMSE performances in both LoS and multi-path cases, before and after refinement. 
This can be partially explained by the low column correlation of the proposed distance-parameterized angular-domain dictionary, as shown in Fig. \ref{Fig:correlation}. According to the compressed sensing theory \cite{Candes_com2008,Yonina_CompressedSensing}, a low column correlation of the dictionary is a primer to guarantee sparse recovery. Therefore, the proposed algorithm enjoys inherent superiority from the dictionary design. 
Furthermore, by comparing the two figures, it is obvious that the
multi-path channel estimation was more challenging as there was about $2\sim 5$ dB performance loss in NMSE for all algorithms employed. This agrees with our previous analysis that interference from other paths will degrade the estimation accuracy. 

\begin{figure}[t]
    \centering
    \includegraphics[width=0.6\textwidth]{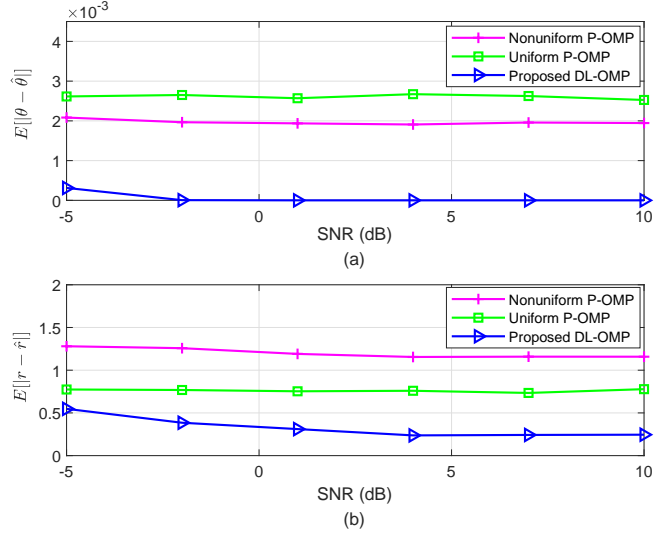}
    \caption{The parameter estimation error against different SNRs. (a) The angle of arrival. (b) The distance. }
    \label{Fig:Sim_3}
\end{figure}

Another observation worth mentioning is that the NMSEs of these estimation algorithms become nearly saturated despite the increase of the SNR. This is because the estimation errors of the angle and/or distance are dominant as compared with the channel noise at high SNRs, as shown in Fig. \ref{Fig:Sim_3}, where the number of pilots is $P=50$ and the users are located in $[5,20]$ m. Note that for the uniform P-OMP, its angle estimation error is comparatively larger than the other two methods, resulting from the high column coherence of the uniform dictionary. While its nonuniform counterpart suffers from significant distance errors. By contrast, both angle and distance estimation errors of the DL-OMP are insignificant. Therefore, accurate user location can also be obtained through DL-OMP based channel estimation.

\begin{figure}[t]
    \centering
    \includegraphics[width=0.6\textwidth]{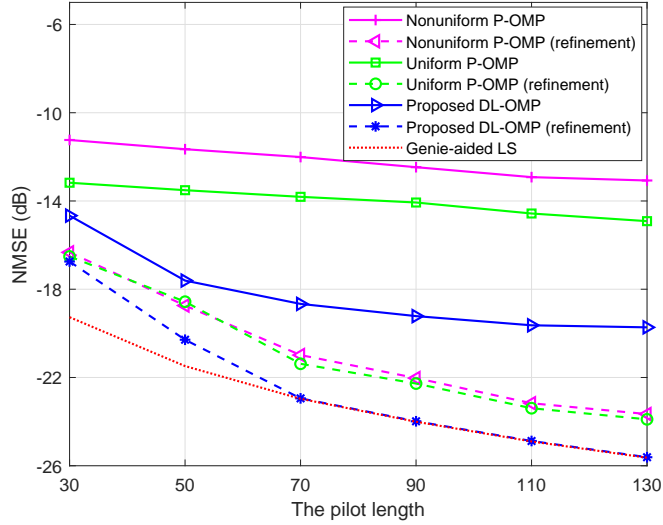}
    \caption{The NMSE of the channel estimate against the length of pilots. }
    \label{Fig:Sim_4}
\end{figure}
\begin{figure}[t]
    \centering
    \includegraphics[width=0.6\textwidth]{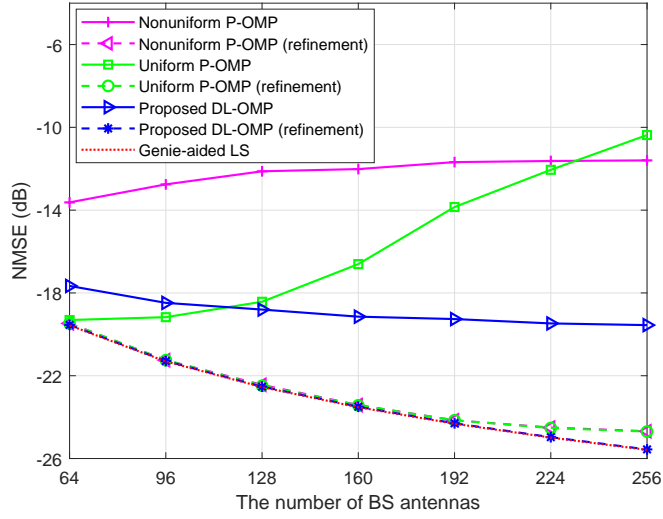}
    \caption{The NMSE of the channel estimate against the number of BS antennas. }
    \label{Fig:Sim_5}
\end{figure}
\begin{figure}[t]
    \centering
    \includegraphics[width=0.6\textwidth]{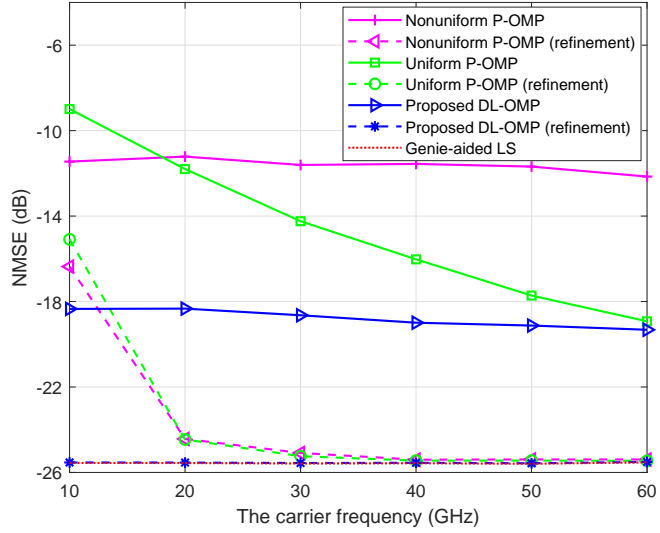}
    \caption{The NMSE of the channel estimate against the carrier frequency. }
    \label{Fig:Sim_6}
\end{figure}
\begin{figure}[t]
    \centering
    \includegraphics[width=0.6\textwidth]{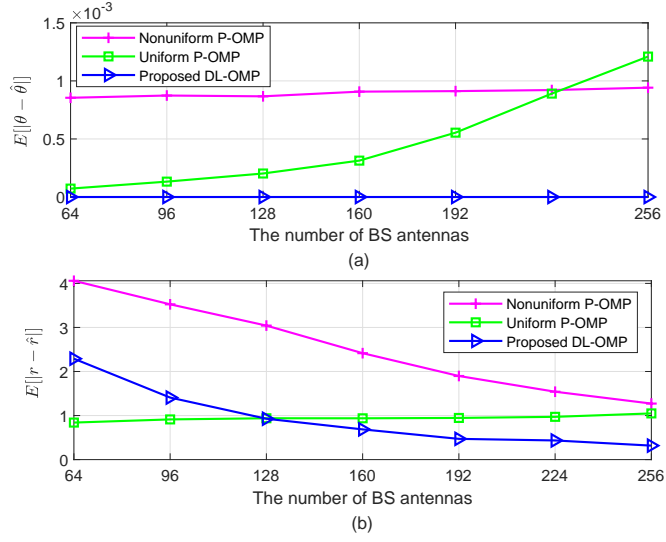}
    \caption{The parameter estimation error against the number of BS antennas. (a) The angle of arrival. (b) The distance. }
    \label{Fig:Sim_7}
\end{figure}
\begin{figure}[t]
    \centering
    \includegraphics[width=0.6\textwidth]{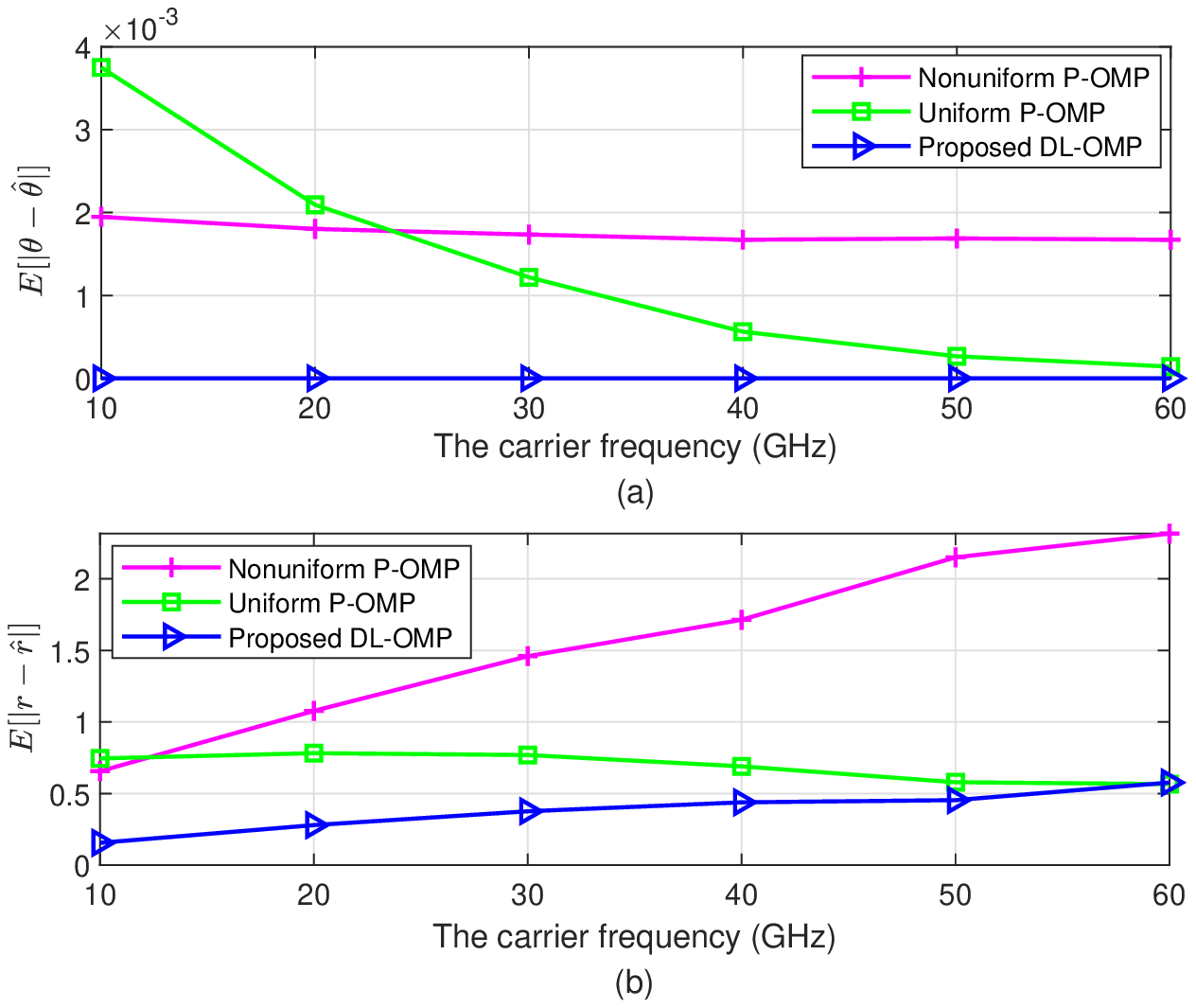}
    \caption{The parameter estimation error against the carrier frequency. (a) The angle of arrival. (b) The distance. }
    \label{Fig:Sim_8}
\end{figure}

To further demonstrate the robustness of the proposed DL-OMP algorithm, we investigated the channel estimation NMSEs with respect to different number of pilots in Fig. 
\ref{Fig:Sim_4}, and the strength of near-field effects in Fig. \ref{Fig:Sim_5} and Fig. \ref{Fig:Sim_6}.
In these figures, the SNR was set as $6$ dB, and other parameters were set as in Table \ref{Tab:II} unless otherwise specified. Here, we only presented the results of the LoS case for brevity.  The results in the multi-path case are similar. In Fig. \ref{Fig:Sim_4}, all curves showed a downtrend with the increase of the pilot length and the superiority of the proposed DL-OMP can be observed over the entire range. However, the comparatively low accuracy over the region of $P<70$ implies that the proposed algorithm works better with more measurements. 
As we mentioned before, the boundary of the near-field, i.e, the Fraunhofer limit, is defined as $d_F=\frac{2D^2}{\lambda}$ with $D$ representing the antenna diameter and $\lambda$ the wavelength. Substituting $D=Nd=N\lambda/2$ and $\lambda=c/f_c$, the Fraunhofer limit can be rewritten as $d_F=\frac{N^2c}{2f_c}$, which 
is proportional to the square of the number of antennas and inversely proportional to the carrier frequency. Therefore, 
in Fig. \ref{Fig:Sim_5} and Fig. \ref{Fig:Sim_6}, we respectively varied the number of BS antennas from $N=64$ to $N=256$, and the carrier frequency from $10$ GHz to $60$ GHz to show the enhancement or weakening of near-field effects in a fixed region $[10,20]$ m.
From the figures, we observe that the proposed algorithm had good robustness and performance advantages in different scenarios. The P-OMPs, however, 
either exhibited significant performance degradation (the uniform one) when the near-field effects became more pronounced, or always maintained at a comparatively poor performance (the nonuniform one). The reason, again, lies in the high column coherence induced angle and distance estimation errors, as supported in Fig. \ref{Fig:Sim_7} and Fig. \ref{Fig:Sim_8}. In particular, with the enhancement of the near-field effects, the uniform P-OMP suffered from higher angle estimation errors, leading to a sharp performance degradation in the channel NMSEs. Therefore, we can speculate that
the uniform method will be even worse than its nonuniform counterpart when the user is very close to the BS, which is in line with the conclusion in \cite{cui2022channel}. By contrast, for the  proposed DL-OMP,  except for its superior angle estimate, its distance estimate became even more accurate with the enhancement of near-field effects, suggesting its appliability in various scenarios.
\section{Conclusions}
\label{sec:Conc}
To date, wireless communications channel estimation algorithms are mainly designed for planar wave based far-field region. However, in the future 6G systems, devices are likely to operate in the radiating near-field region with the usage of large-scale antennas and high frequencies. Therefore, both the channel representation and the corresponding estimation algorithms need to be reconsidered. In this work, we have mitigated the gap by first proposing a distance-parameterized angular-domain sparse model to represent the near-field channel. The proposed representation distinguishes itself from the polar-domain two dimensional method by making the distance an unknown parameter of the dictionary, thus the dictionary size only depends on the angular resolution. This benefits the proposed method with low storage and dictionary coherence from the dictionary design stage. Subsequently, two dictionary learning orthogonal matching pursuit (DL-OMP) algorithms have been designed respectively for the line of sight and multi-path channels, followed by restricted isometry property (RIP) based recovery condition analysis and the storage and complexity analyses. These theoretical analyses, together with simulations in multi-user communications, further verified the effectiveness and superiority of the proposed model and estimation algorithms. 
\begin{appendices}
\section{Proof of \textit{Proposition 1}}
\label{sec:App_B}
\textit{Proof:}
From the definition of the column correlation, we have
\begin{eqnarray}
\epsilon (\theta_p,\theta_q,r_p,r_q)&=&\left| \frac{1}{N}\textbf{b}_N^H(\theta_p,r_p)\textbf{b}_N(\theta_q,r_q)\right|
\nonumber\\
&=& \left|\frac{1}{N} \sum_{n=1}^{N} e^{j2\pi \frac{f_c}{c} (r_p^{(n)}-r_q^{(n)}-r_p^{(1)}+r_q^{(1)})}\right|,
\label{Eq:ep_1}
\end{eqnarray}
where $r_p^{(n)}$ and $r_q^{(n)}$ are defined by \eqref{Eq:rn}. 
By substituting \eqref{Eq:rn} into \eqref{Eq:ep_1} and after a few mathematical manipulations, we have
\begin{eqnarray}
\epsilon (\theta_p,\theta_q,r_p,r_q)=\left|\frac{1}{N} \sum_{n=1}^{N} e^{j2\pi \frac{f_c}{c} \left(
\left(r_p^{(1)}-r_q^{(1)}\right)\left(\frac{r_p^{(1)}+r_q^{(1)}}{r_p^{(n)}+r_q^{(n)}}-1\right)-2(n-1)d\frac{r_p^{(1)}\text{cos}\theta_p-r_q^{(1)}\text{cos}\theta_q}{r_p^{(n)}+r_q^{(n)}}
\right)}\right|.
\label{Eq:ep_2}
\end{eqnarray}
The coherence of the dictionary is defined as the largest correlation between any two columns, that is,  $\epsilon=\max\limits_{p\ne q}\epsilon (\theta_p,\theta_q,r_p,r_q)$ \cite{tropp2007signal}. For $\epsilon (\theta_p,\theta_q,r_p,r_q)$ given in \eqref{Eq:ep_2}, the maximum may be achieved when the two columns share one of the parameter (angle or distance) while differ in the other parameter with only one sampling interval.
Therefore, we here consider two extreme cases: $r_p^{(1)}=r_q^{(1)}$  and $\theta_p=\theta_q$.
For $\textbf{W}(\textbf{r})$, since each column is assigned a different angle, only $r_p^{(1)}=r_q^{(1)}$ may occur. While for the polar-domain 2D dictionary, both cases exist.

1) 
When $r_p^{(1)}=r_q^{(1)}$, we have
\begin{eqnarray}
\epsilon (\theta_p,\theta_q,r_p,r_q)=\left|\frac{1}{N} \sum_{n=1}^{N} e^{j2\pi \frac{f_c}{c} \left(
-2(n-1)d\left(\text{cos}\theta_p-\text{cos}\theta_q\right)\frac{r_p^{(1)}}{r_p^{(n)}+r_q^{(n)}}
\right)}\right|.
\label{Eq:ep_3}
\end{eqnarray}
As the angle space $\text{cos}\theta\in[-1,1]$ is assumed to be divided into $N$ parts, the minimum angular sampling interval is set as 
$\text{cos}\theta_p-\text{cos}\theta_q=\frac{2}{N}$. 
Meanwhile, by employing the following approximation \cite{sherman1962properties,cui2022channel},
\begin{equation}
r^{(n)}\approx r^{(1)}-(n-1)d\text{cos}\theta+\frac{(n-1)^2d^2(1-\text{cos}^2\theta)}{2r^{(1)}},
\label{Eq:approx_rneq}
\end{equation}
and define $z=(\text{cos}\theta_p+\text{cos}\theta_q)/r_p^{(1)}$,
we have 
%
\begin{equation}
\frac{r_p^{(1)}}{r_p^{(n)}+r_q^{(n)}}\approx \frac{1}{2-(n-1)dz+(n-1)^2d^2z}.
\end{equation}
Then, \eqref{Eq:ep_3} can be rewritten as
\begin{eqnarray}
\epsilon (\theta_p,\theta_q,r_p,r_q)
\approx \left|\frac{1}{N} \sum_{n=1}^{N} e^{j2\pi \frac{f_c}{c} \left(
-\frac{4(n-1)d}{N\left(2-(n-1)dz+(n-1)^2d^2z\right)}\right)}
\right|.
\end{eqnarray}

2) When $\theta_p=\theta_q$,
by setting the distance sampling interval as $1$ meter, i.e., $r_p^{(1)}-r_q^{(1)}=1$, we have
\begin{eqnarray}
\epsilon (\theta_p,\theta_q,r_p,r_q)=\left|\frac{1}{N} \sum_{n=1}^{N} e^{j2\pi \frac{f_c}{c} \left(
\frac{r_p^{(1)}+r_q^{(1)}-2(n-1)d\text{cos}\theta_p}{r_p^{(n)}+r_q^{(n)}}-1
\right)}\right|.
\label{Eq:ep_3_2}
\end{eqnarray}
Again by employing \eqref{Eq:approx_rneq}, we can re-express
\eqref{Eq:ep_3_2} as
\begin{eqnarray}
\epsilon (\theta_p,\theta_q,r_p,r_q)\approx\left|\frac{1}{N} \sum_{n=1}^{N} e^{j2\pi \frac{f_c}{c} \left(
\frac{1}{1+(n-1)^2d^2 x}-1
\right)}
\right|,
\label{Eq:ep_4}
\end{eqnarray}
were $x=\frac{1-\text{cos}^2\theta_p}{2r_p^{(1)}r_q^{(1)}}$.

For dictionary $\textbf{W}(\textbf{r})$, only case 1) may occur, therefore, we have the dictionary coherence, denoted as $\epsilon_{\textbf{W}}(\theta_p,\theta_q,r_p)$, expressed in \eqref{Eq:p1_epW}. While for dictionary $\textbf{D}$, both cases exist, the correlation in case 2) is generally much larger. Therefore, the dictionary coherence $\epsilon_{\textbf{D}}(\theta_p,r_p,r_q)$ is obtained as \eqref{Eq:p1_epD}.
This completes the proof of \textit{Proposition 1}.
\section{Proof of \textit{Proposition 2}}
\label{App_C}
\textit{Proof:}
The updating of the dictionary aims to mitigate the distance mismatch between atoms in the dictionary and the received signal. 
Mathematically, the degree of matching is measured by the absolute inner product, e.g., 
the absolute inner product between the selected atom $\textbf{b}_P(\hat{\theta}_1,r_{\text{max}})$ of $\textbf{W}_1^{\circ}$ and the received signal $\textbf{y}_1$ is given by
\begin{eqnarray}
\xi=\left|\textbf{b}^H_P(\hat{\theta}_1,r_{\text{max}})\textbf{y}_1\right|& = &\left|g e^{-j2\pi \frac{f_c}{c}r_1}\sum_{p=1}^{P/2}e^{-j2\pi \frac{f_c}{c}(r_1^{(p)}-r_1-r_{\text{max}}^{(p)}+r_{\text{max}})}\right|\nonumber
\\
&\approx&
\left|g e^{-j2\pi \frac{f_c}{c}r_1}\sum_{p=1}^{P/2}e^{-j2\pi \frac{f_c}{c}((p-1)^2d^2(1-\text{cos}^2\hat{\theta}_1))(\frac{1}{2r_1}-\frac{1}{2r_\text{max}})}\right|,
\label{Eq:corr_los_1}
\end{eqnarray}
where the approximation in the second step is obtained by assuming
\begin{eqnarray}
r_1^{(p)}-r_1&=&\sqrt{ r_1^2+(p-1)^2d^2-2(p-1)d r_1 \text{cos}\theta_1}-r_1\nonumber\\
&\approx & -(p-1)d\text{cos}\theta_1+\frac{(p-1)^2d^2(1-\text{cos}^2\theta_1)}{2r_1},
\\
r_{\text{max}}^{(p)}-r_{\text{max}}&\approx & -(p-1)d\text{cos}\hat{\theta}_1+\frac{(p-1)^2d^2(1-\text{cos}^2\hat{\theta}_1)}{2r_{\text{max}}},
\end{eqnarray}
%
%
and $\theta_1\approx \hat{\theta}_1$. 

From \eqref{Eq:corr_los_1}, we see that the mismatch between the selected atom and the received signal can be eliminated by modifying the $p$th element of $\textbf{b}_P(\hat{\theta}_1,r_{\text{max}})$ with $e^{-j2\pi \frac{f_c}{c}((p-1)^2d^2(1-\text{cos}^2\hat{\theta}_1))(\frac{1}{2r_1}-\frac{1}{2r_\text{max}})}$.
To this end, after obtaining a new distance estimation, in steps 6 and 7 of \textit{Algorithm 1}, we update the dictionary $\textbf{W}_1$ in a similar way as
\begin{equation}
\textbf{W}_1=\text{diag}(\boldsymbol\alpha_1)\textbf{W}^{\circ}_1,
\end{equation}
where
\begin{equation}
\boldsymbol\alpha_1=\left[1, e^{-j2\pi \frac{f_c}{c}(d^2(1-\text{cos}^2\hat{\theta}_1))(\frac{1}{2\hat{r}_1}-\frac{1}{2r_{\text{max}}})},\cdots, e^{-j2\pi \frac{f_c}{c}((P/2-1)^2d^2(1-\text{cos}^2\hat{\theta}_1))(\frac{1}{2\hat{r}_1}-\frac{1}{2r_{\text{max}}})}\right]^T.
\end{equation}
Similar updating can be performed for $\textbf{W}_2$.
This completes the proof of \textit{Proposition 2}.
\section{Proof of \textit{Theorem 1}}
\label{sec:App_A}
\textit{Proof:} To prove \textit{Theorem 1}, we will demonstrate that when \eqref{Eq:bound} holds, the support set of $\textbf{s}$ will be successfully recovered in $L$ iterations.
To begin with, consider the first iteration where $\Lambda^0=\varnothing$.
With the dictionary $\textbf{W}(\hat{\textbf{r}})$ at hand, the matching step yields
\begin{equation}
\boldsymbol{\psi}^1=\textbf{W}^H(\hat{\textbf{r}}) \textbf{y}= \textbf{W}^H(\hat{\textbf{r}})(\textbf{W}(\hat{\textbf{r}})\textbf{s}+\textbf{e}),
\end{equation}
of which the $i$th element is
\begin{equation}
\boldsymbol{\psi}^1(i)= <\textbf{W}(\hat{\textbf{r}})\boldsymbol\zeta_i,\textbf{W}(\hat{\textbf{r}})\textbf{s}+\textbf{e}>,
\end{equation}
where $\boldsymbol\zeta_i$ denotes the $i$th natural basis.
Define
\begin{equation}
{\psi}^1_{\text{max}}= \max\limits_{i\in \text{supp}(\textbf{s})}|\boldsymbol{\psi}^1(i)|,
\end{equation}
and $\textbf{U}=|<\textbf{W}(\hat{\textbf{r}})\textbf{s},\textbf{W}(\hat{\textbf{r}})\textbf{s}+\textbf{e}>|$. On one hand,
\begin{equation}
\textbf{U}=\left|\sum\textbf{s}(i)\boldsymbol{\psi}^1(i)\right|\le ||\textbf{s}||_1 {\psi}^1_{\text{max}}\le \sqrt{L}||\textbf{s}||_2 {\psi}^1_{\text{max}}.
\end{equation}
On the other hand,
\begin{equation}
\textbf{U} \ge ||\textbf{W}(\hat{\textbf{r}})\textbf{s}||_2^2 - ||\textbf{W}(\hat{\textbf{r}})\textbf{s}||_2 ||\textbf{e}||_2\ge (1-\delta_{L+1})||\textbf{s}||_2^2 - \sqrt{1+\delta_{L+1}} ||\textbf{s}||_2||\textbf{e}||_2.
\label{Eq:ulb}
\end{equation}
Therefore, we have
\begin{equation}
{\psi}^1_{\text{max}}\ge \frac{1}{\sqrt{L}}\left( (1-\delta_{L+1})||\textbf{s}||_2- \sqrt{1+\delta_{L+1}}||\textbf{e}||_2\right).
\label{Eq:psi_max}
\end{equation}
For $i\notin \text{supp}(\textbf{s})$, according to \cite{Candes_com2008}, we have
\begin{eqnarray}
|\boldsymbol{\psi}^1(i)|&=&\left|<\textbf{W}(\hat{\textbf{r}})\boldsymbol\zeta_i,\textbf{W}(\hat{\textbf{r}})\textbf{s}>
+<\textbf{W}(\hat{\textbf{r}})\boldsymbol\zeta_i,\textbf{e}>\right|
\nonumber\\
&\le& \delta_{L+1} ||\textbf{s}||_2 + ||\textbf{W}(\hat{\textbf{r}})\boldsymbol\zeta_i||_2 ||\textbf{e}||_2
\nonumber\\
&\le& \delta_{L+1} ||\textbf{s}||_2 + \sqrt{1+\delta_{L+1}}||\textbf{e}||_2.
\label{Eq:psi_min}
\end{eqnarray}
When \eqref{Eq:bound} hold, it is straightforward to verify that
\begin{eqnarray}
{\psi}^1_{\text{max}} > |\boldsymbol{\psi}^1(i)|,\quad \forall i\notin \text{supp}(\textbf{s}),
\end{eqnarray}
which guarantees the success of the first iteration.

Then, in the $l$th iteration, suppose that all previous iterations succeed, which implies that $\Lambda^{l-1}$ is a subset of $\text{supp}(\textbf{s})$.
Define $\tilde{\textbf{s}}=\textbf{s}-\textbf{s}|_{\Lambda^{l-1}}$, and $\tilde{\textbf{e}}=-(\Delta \textbf{W})\tilde{\textbf{s}}+\textbf{v}$, then $\text{supp}(\tilde{\textbf{s}})\subseteq \text{supp}(\textbf{s})$. The mapping result is given by
\begin{equation}
\boldsymbol{\psi}^{l}=\textbf{W}^H(\hat{\textbf{r}})(\textbf{W}(\hat{\textbf{r}})\tilde{\textbf{s}}+\tilde{\textbf{e}}).
\end{equation}
Define
$\boldsymbol{\psi}^l(i)= <\textbf{W}(\hat{\textbf{r}})\boldsymbol\zeta_i,\textbf{W}(\hat{\textbf{r}})\tilde{\textbf{s}}+\tilde{\textbf{e}}>$
and $\textbf{U}=|<\textbf{W}(\hat{\textbf{r}})\tilde{\textbf{s}},\textbf{W}(\hat{\textbf{r}})\tilde{\textbf{s}}+\tilde{\textbf{e}}>|$,
it can be derived that
\begin{equation}
\textbf{U}=\left|\sum\tilde{\textbf{s}}(i)\boldsymbol{\psi}^l(i)\right|\le ||\tilde{\textbf{s}}||_1 {\psi}^l_{\text{max}}\le \sqrt{L^{\prime}}||\tilde{\textbf{s}}||_2 {\psi}^l_{\text{max}},
\end{equation}
where $L^{\prime}=L-l+1$. Following similar steps in the proof for the first iteration, we have
%
\begin{equation}
{\psi}^l_{\text{max}}\ge \frac{1}{\sqrt{L^{\prime}}}\left( (1-\delta_{L^{\prime}+1})||\tilde{\textbf{s}}||_2- \sqrt{1+\delta_{L^{\prime}+1}}||\tilde{\textbf{e}}||_2\right)\ge \frac{1}{\sqrt{L}}\left( (1-\delta_{L+1})||\tilde{\textbf{s}}||_2- \sqrt{1+\delta_{L+1}}||\tilde{\textbf{e}}||_2\right),
\end{equation}
where the second inequality holds since according to the properties of RIP, we have $\delta_{L^{\prime}+1}<\delta_{L+1}$.
Similarly,
\begin{eqnarray}
|\boldsymbol{\psi}^l(i)|\le \delta_{L^{\prime}+1} ||\tilde{\textbf{s}}||_2 + \sqrt{1+\delta_{L^{\prime}+1}}||\tilde{\textbf{e}}||_2
\le
\delta_{L+1} ||\tilde{\textbf{s}}||_2 + \sqrt{1+\delta_{L+1}}||\tilde{\textbf{e}}||_2
,  \quad i\notin \text{supp}(\tilde{\textbf{s}}).
\end{eqnarray}
Since we approximately have $\frac{||\tilde{\textbf{e}}||^2_2}{||\tilde{\textbf{s}}||^2_2}\approx \frac{||\textbf{e}||^2_2}{||\textbf{s}||^2_2}=\mu$, the condition given by \eqref{Eq:bound} also guarantees the success of the $l$th iteration, that is, ${\psi}^l_{\text{max}} > |\boldsymbol{\psi}^l(i)|, \forall i\notin \text{supp}(\tilde{\textbf{s}})$.
This completes the proof of \textit{Theorem 1}.
%
\end{appendices}
\bibliographystyle{IEEEtran}
\bibliography{IEEEabrv,yourbibliography}		
\end{document}